\documentclass[a4paper,11pt,article,twoside]{memoir}

\pdfoutput=1

\usepackage[utf8]{inputenc} 
\usepackage{graphicx}
\usepackage{titlesec}
\usepackage[resetlabels]{multibib}
\usepackage{verbatim}
\usepackage{siunitx}
\usepackage{units}
\usepackage{placeins}
\usepackage{amsmath}
\usepackage[linktocpage,hypertexnames=false]{hyperref}

\newcommand{\Xmax}{$X_\textrm{max}$\xspace}
\newcommand{\Xmaxmath}{X_\textrm{max}} 

\def\figref#1{Fig.~\ref{fig:#1}}
\def\figlab#1{\label{fig:#1}}  
\def\tabref#1{Table~\ref{tab:#1}}
\def\tablab#1{\label{tab:#1}}  
\def\eqref#1{Eq.~\ref{eq:#1}}

\def\eqlab#1{\label{eq:#1}}
\newcommand*{\secref}[1]{Section~\ref{sec:#1}}
\newcommand*{\seclab}[1]{\label{sec:#1}}

\def\vB{{\mathbf{v}\!\times\!\mathbf{B}}}
\def\vvB{{\mathbf{v}\!\times\!\left(\vB\right)}}
\def\vz{{\mathbf{v}\!\times\!\mathbf{z}}}
\def\vvz{{\mathbf{v}\!\times\!\left(\vz\right)}}

\makeatletter

\DeclareRobustCommand\xspace{\@xspace@firsttrue
	\futurelet\@let@token\@xspace}

\newif\if@xspace@first
\def\@xspace@simple{\futurelet\@let@token\@xspace}

\let\@xspace@eTeX@setup\relax

\def\@xspace@exceptions@tlp{%
	,.’/?;:!~-)\ \/\bgroup\egroup\@sptoken\space\@xobeysp
	\footnote\footnotemark
	\xspace@check@icr
}

\def\@xspace@break@loop#1\@nil{}

\def\@xspace{%
	\@xspace@lettoken@if@letter@TF \space{%
		\if@xspace@first
			\@xspace@firstfalse
			\let\@xspace@maybespace\space
			\@xspace@eTeX@setup
		\fi
		\expandafter\@xspace@check@token
			\@xspace@exceptions@tlp\@xspace@q@nil\@nil
		\@xspace@token@if@equal@NNT \space \@xspace@maybespace
		{%
			\@xspace@lettoken@if@expandable@TF
			{\expandafter\@xspace@simple}%
			{\@xspace@maybespace\@xspace@hook}%
		}%
	}%
}

\def\@xspace@check@token #1{%
	\ifx\@xspace@q@nil#1%
		\expandafter\@xspace@break@loop
	\fi
	\expandafter\ifx\csname @let@token\endcsname#1%
		\let\@xspace@maybespace\relax
		\expandafter\@xspace@break@loop
	\fi
	\@xspace@check@token
}

\def\@xspace@lettoken@if@letter@TF{%
	\ifcat\noexpand\@let@token @
		\expandafter\@firstoftwo
	\else
		\expandafter\@secondoftwo
	\fi
}
\def\@xspace@lettoken@if@expandable@TF{%
	\expandafter\ifx\noexpand\@let@token\@let@token%
		\expandafter\@secondoftwo
	\else
		\expandafter\@firstoftwo
	\fi
}
\def\@xspace@token@if@equal@NNT#1#2{%
	\ifx#1#2%
		\expandafter\@firstofone
	\else
		\expandafter\@gobble
	\fi
}

\def\ps@PoS{%
\def\@oddhead{%
\smash{%
\vbox{\hsize=\textwidth\noindent%
\if@shorttitle\copy\PoSshorttitle@box\else{\small\it\@title}\fi\hfill
			\copy\@firstaubox\vskip.17em \hrule
}}
}%
\let\@mkboth\@gobbletwo  
\let\sectionmark\@gobble
\let\subsectionmark\@gobble
}

\def\logo{\raisebox{5mm}{\hbox{\noindent\includegraphics[height=23mm]{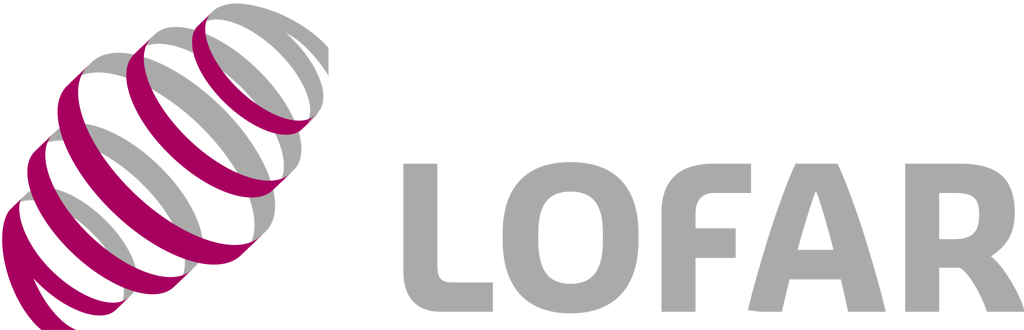}}}}

\newskip\titlesk@a		
\newskip\titlesk@b		%
\newskip\titlesk@c		%
\newif\if@titlepage\@titlepagetrue
\def\@maketitle{
  \begin{flushleft}
    \hspace{\fill}\logo
    \vskip \titlesk@b		   
    {\huge \sffamily              
      \bfseries
      \@title\par}%
    \vskip \titlesk@b              
    \hrule height 1\p@
    \vskip \titlesk@b		   
    {\normalsize \bfseries \sffamily
      \@author \par}
    \vskip \titlesk@c              
    \centerline{\parbox{.9\textwidth}
      {\abstract@cs}}%
    \dedic@ted                     
    \@PACS                         
    \vskip \titlesk@c
    \vfill
    \FullConference@box            
  \end{flushleft}%
  \@titlepagefalse
}
\newcommand\@auth{\parskip\z@
		      \def\\{\egroup			
			\par				
			\vskip\addr@skip		
			\vbox\bgroup\hsize=.9\textwidth	%
			\normalfont\itshape\small\raggedright		
			\def\\{\par\leavevmode\ignorespaces}
			\leavevmode\ignorespaces}%
			}
\newtoks\prev@t
\newtoks\cur@t
\newbox\@firstaubox
\newskip\addr@skip\addr@skip=.38em
\newskip\auth@skip\auth@skip=.6em\@plus.03fil\@minus.2ex

\renewcommand\author[1]{%
		\prev@t=\expandafter{\@auth}
		\cur@t={\vskip\auth@skip		
			\vtop\bgroup\large #1\egroup		
			\par}
		\long\xdef\@author{\the\prev@t\the\cur@t}
		}
\newcommand\addauthor[1]{%
		\prev@t=\expandafter{\@author}
		\cur@t={\vskip\auth@skip		
			\vtop\bgroup\large #1\egroup		
			\par}
		\long\xdef\@author{\the\prev@t\the\cur@t}
		}
\renewcommand{\abstract}[1]
	{\gdef\abstract@cs{		
	{\small
	  \ignorespaces #1}}
	\normalsize}
\newif\if@leadauthor

\newcommand\speaker[1]{
			\setcounter{footnote}{0}
			\if@leadauthor
				\@leadauthorfalse%
			\else	
				\global\setbox\@firstaubox
				\hbox{{\let\thanks\@gobble
					\let\footnote\@gobble
					\small\rmfamily #1}}%
			\fi
			#1\thanks{Speaker.}\
			}
\newcommand\leadauthor[1]{
			\global\setbox\@firstaubox
			\hbox{{\let\thanks\@gobble
				\let\footnote\@gobble
				\small\rmfamily #1}}%
			#1\@leadauthortrue
			}

\newif\if@shorttitle
\newbox\PoSshorttitle@box
\newcommand\ShortTitle[1]{\global\setbox\PoSshorttitle@box\hbox{\small\textit{#1}}\@shorttitletrue}
\gdef\FullConference@box{\relax}
\newcommand\FullConference[1]{\gdef\FullConference@box{
			 \vbox{\small\raggedright
			   \textit{#1}}}}
\newcommand\email[1]{{\href{mailto:#1}{\ttfamily\upshape #1}}}
\gdef\dedic@ted{\relax}
\newcommand\dedicated[1]{\gdef\dedic@ted{\vskip .5\titlesk@c
                              \vbox{\small\raggedleft\textit{#1}}}}
\gdef\@PACS{\relax}
\newcommand\PACS[1]{\gdef\@PACS{\vskip.5\titlesk@c
                                \vbox{\small\ttfamily\raggedleft PACS:\ #1}}}
\gdef\PoSspecial@url{\relax}
\newcommand\PoSspecialurl[1]{\@PoSspecialurltrue\gdef\PoSspecial@url{%
    \hbox{\tiny\ttfamily #1}}}
\long\def\@makecaption#1#2{%
  \vskip\abovecaptionskip
 {\let\label\@gobble
  \let\index\@gobble
  \let\glossary\@gobble
  \sbox\@tempboxa{\small {\bfseries #1:} #2}
  \global\dimen0\wd\@tempboxa}
  \ifdim \dimen0 >\hsize
    \small {\bfseries #1:} #2\par
  \else
    \global\@minipagefalse \sbox\@tempboxa{\small {\bfseries #1:} #2}%
    \hb@xt@\hsize{\hfil\box\@tempboxa\hfil}%
  \fi
  \vskip\belowcaptionskip}

\renewcommand{\part}[1]{%
	\addtocounter{part}{1}
	\setcounter{chapter}{0}
	\setcounter{equation}{0}
	\setcounter{figure}{0}
	\setcounter{table}{0}
	\global\def\@thanks{}
	\addtocontents{toc}{\vspace{4.2ex}\protect\contentsline{chapter}{\protect\numberline{}\large\hspace{-1.67em} #1}{}{}}
	\ShortTitle{#1}
	\FloatBarrier}

\makeatother

\newcommand{\nocontentsline}[3]{}
\newenvironment{tocless}{\bgroup\let\addcontentsline\nocontentsline}{\egroup}

\renewcommand{\it}{\itshape}

\titleformat*{\section}{\large\bfseries}

\let\subsection\section
\let\section\chapter

\newcites{lora}{{\large\bfseries References}}
\newcites{escale}{{\large\bfseries References}}
\newcites{gdas}{{\large\bfseries References}}
\newcites{xmax}{{\large\bfseries References}}
\newcites{efields}{{\large\bfseries References}}

\title{Contributions of the LOFAR Cosmic Ray\\ Key Science Project to the 36th\,International Cosmic Ray Conference (ICRC 2019)}

\ShortTitle{Contributions of the LOFAR Cosmic Ray Key Science Project to the ICRC 2019}

\author{A.~Bonardi$^{1}$, S.\,Buitink$^{2,1}$, A.\,Corstanje$^{1,2}$, H.\,Falcke$^{1,3,4}$, B.\,M.\,Hare$^{5}$,
J.\,R.\,H\"{o}randel$^{1,2,3}$, T.\,Huege$^{6,2}$, G.\,Krampah$^{2}$, P.\,Mitra$^{2}$,
K.\,Mulrey$^{2}$, A.\,Nelles$^{7,8}$, H.\,Pandya$^{2}$, J.\,P.\,Rachen$^{2,1}$, L.\,Rossetto$^{1}$,
O.\,Scholten$^{5,9}$, S.\,ter\,Veen$^{4}$, T.\,N.\,G.\,Trinh$^{5,10}$, T.\,Winchen$^{2}$\\
\\
$^1$ Department of Astrophysics/IMAPP, Radboud University, P.O. Box 9010, 6500 GL Nijmegen, The Netherlands\\
$^2$ Astrophysical Institute, Vrije Universiteit Brussel, Pleinlaan 2, 1050\,Elsene, Belgium\\
$^3$ NIKHEF, Science Park Amsterdam, 1098 XG Amsterdam, The Netherlands\\
$^4$ Netherlands Institute of Radio Astronomy (ASTRON), Postbus 2, 7990 AA Dwingeloo,\\ The Netherlands\\
$^5$ KVI-CART, University Groningen, P.O. Box 72, 9700 AB Groningen\\
$^6$ Institut f\"{u}r Kernphysik, Karlsruhe Institute of Technology (KIT), P.O.\,Box 3640, 76021\,Karlsruhe, Germany\\
$^7$ DESY, Platanenallee 6, 15738 Zeuthen, Germany\\
$^8$ ECAP, Friedrich-Alexander-University Erlangen-N\"{u}rnberg, 91058 Erlangen, Germany\\
$^9$ Interuniversity Institute for High-Energy (IIHE), Vrije Universiteit Brussel, Pleinlaan 2, 1050\,Elsene, Belgium\\
$^{10}$Department of Physics, School of Education, Can Tho University Campus II, 3/2 Street,\\ Ninh Kieu District, Can Tho City, Vietnam\\
}

\abstract{This is a collection of papers that have been contributed by the LOFAR Cosmic Ray Key Science Project (CRKSP) to the 36th International Cosmic Ray Conference held in Madison, Wisconsin, on July 24th to August 1st, 2019 (ICRC 2019). All papers contained here have been individually published in PoS(ICRC2019) with paper numbers 205, 362, 352, 416, and 363, in the order they appear in this collection. Minor modifications to the PoS versions have been applied where appropriate.} 

\FullConference{\ }

\settypeblocksize{0.7\paperheight}{0.72\paperwidth}{*}
\setlrmargins{*}{*}{0.67}
\checkandfixthelayout

\begin{document}

\titlingpageend{\clearpage\vspace*{\fill}\noindent Edited by J\"org P. Rachen \thispagestyle{empty}\newpage}

\begin{titlingpage}
\maketitle
\end{titlingpage}

\titlingpageend{\clearpage}

\setcounter{page}{1}

\begin{small}
\tableofcontents*
\end{small}

\vspace{\fill}

\begin{footnotesize}

\subsection*{Acknowledgements\footnotemark}

The LOFAR Cosmic-Ray Key Science Project acknowledges funding from an Advanced Grant of the European Research Council (FP/2007-2013) / ERC Grant Agreement No 227610. The project has also received funding from the European Research Council (ERC) under the European Union’s Horizon 2020 research and innovation programme (grant agreement No 640130). We furthermore acknowledge financial support from FOM (FOM-project 12PR304).  AN  and TW have been supported by DFG grants NE 2031/2-1 and WI 4946/1-1, respectively. LOFAR, the Low Frequency Array designed and constructed by ASTRON, has facilities in several countries, that are owned by various parties (each with their own funding sources), and that are collectively operated by the International LOFAR Telescope foundation under a joint scientific policy.

\footnotetext{%
These ackowledgements are extended by the LOFAR Cosmic Ray Key Science Project in general and apply to all research presented here. Acknowledgments for specific research projects, if any, are given in the individual papers.}

\newpage

\end{footnotesize}


\renewcommand{\logo}{\vspace*{23mm}}

\FullConference{36th International Cosmic Ray Conference -ICRC2019-\\
		July 24th - August 1st, 2019\\
		Madison, WI, U.S.A.}


\savepagenumber
\part{Towards an improved mass composition analysis with LOFAR}

\title{Towards an improved mass composition analysis with LOFAR}

\author{\leadauthor{A. Corstanje},\speaker{S. Buitink}\! for the LOFAR CRKSP\\
		E-mail: \email{a.corstanje@astro.ru.nl}}

\abstract{The LOFAR radio telescope measures air showers in the energy range $10^{17}$ to $\unit[10^{18}]{eV}$.
For each measured shower, the depth of shower maximum \Xmax is reconstructed by simulating the radio signal for an ensemble of showers using CORSIKA and CoREAS. 
Fitting their radio `footprints' on the ground to the measured radio data yields an \Xmax estimate to a precision of about $\unit[20]{g/cm^2}$.
Compared to previous works, we have improved the method in several ways.
Local atmospheric data and refractive index profiles are now included into the simulations.
The energy estimate and the fitting procedure are now done using the radio signals only, thus limiting systematic uncertainties due to the particle detector array (LORA).
Using selection criteria from a more elaborate characterisation of the radio and particle detection, we reduce a composition bias in the \Xmax reconstruction.
A possible residual bias has been bounded from above.
Thus, the systematic uncertainties on $\left<\Xmaxmath\right>$ have been lowered, reducing an important limiting factor for composition studies at any level of statistics. 
} 

\begin{titlingpage}
\maketitle
\end{titlingpage}
\restorepagenumber
\addtocounter{page}{1} 

\section{Introduction}
LOFAR is the most densely instrumented radio-based cosmic-ray observatory in the world. In the center of the array, air showers are detected by hundreds of antennas simultaneously, probing the intensity, shape and polarization of the radio footprint at the ground. These high-resolution measurements have played an important role in the understanding and verification of the radio emission mechanism of air showers. It also allows for reconstruction of the depth of shower maximum, \Xmax, with a resolution of the order of 20 g/cm$^2$, comparable to the fluorescence detection technique. We have previously demonstrated that LOFAR is thus capable of studying mass composition in the energy range of ${\sim}\, 10^{17}{-}10^{18}$ eV. Since then, various improvements have been made to the antenna calibration, the accuracy of simulations and the reconstruction techniques. Here, we highlight the most important changes and discuss the achieved reduction in systematic uncertainties. 

\section{Method}\label{sect:method}
We reconstruct \Xmax of each measured air shower that was detected by at least 3 LOFAR stations. The reconstruction is based on the procedure in \cite{xmax:Buitink:2014}.
To this end, we first simulate 600 showers with CONEX \cite{xmax:Bergmann:2007}, which is very fast compared to a full CoREAS simulation. 
From this set, a subset of about 30 showers are selected which span the natural range of \Xmax. Ten of these are chosen in an interval $\pm \unit[20]{g/cm^2}$ from a first estimate of \Xmax based on a parametrization of the radio footprint \cite{xmax:Nelles_param:2015}, to have a denser sampling there.
These showers are fully simulated with CORSIKA \cite{xmax:Corsika:1998} and CoREAS \cite{xmax:Huege:2013}

\begin{figure}[t]
\begin{center}
\includegraphics[trim={3cm 0 3cm 0}, width=1.023\textwidth]{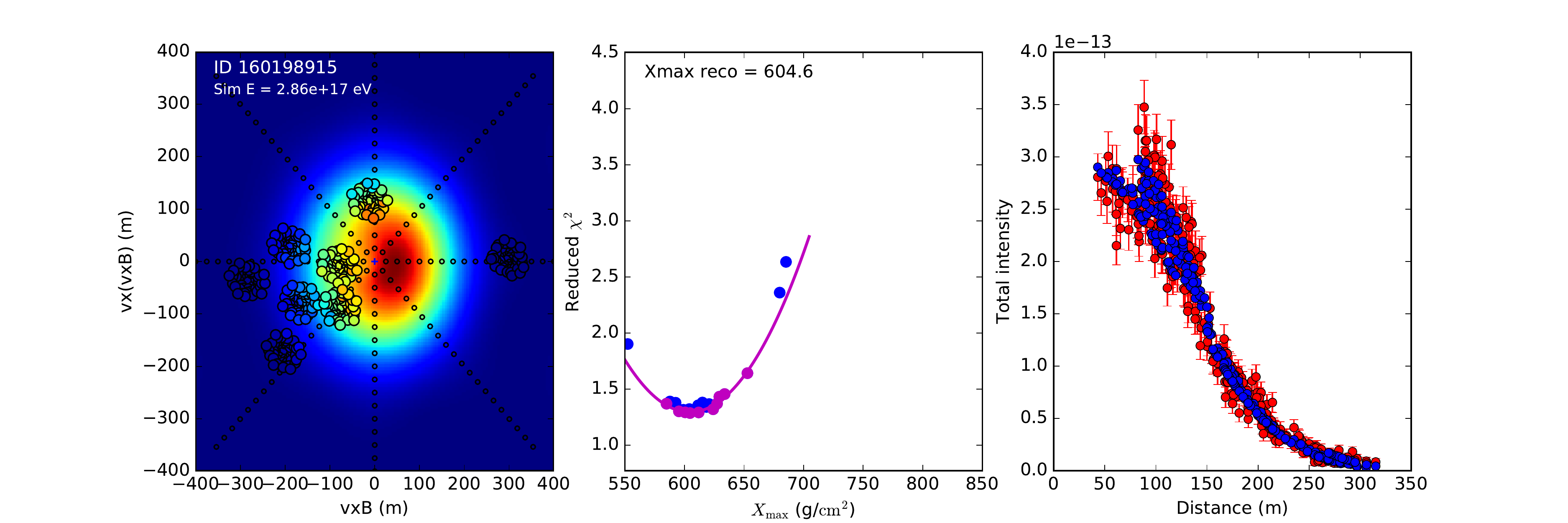}
\caption{\textbf{Left panel:} Signal intensity for best-fitting shower, with LOFAR measurements, in the shower plane. \textbf{Middle panel:} Reduced $\chi^2$ as function of \Xmax, with parabolic fit through the lower envelope of points. \textbf{Right panel:} Projected 1-D lateral distribution, with the measurements as red points with margins, and the simulation values as blue points.}
\label{fig:event_fit_example}
\end{center}
\end{figure}

Their radio signal intensities are fitted to the measured LOFAR data, having the shower core position and an overall scale factor as free parameters. 
This gives a $\chi^2$-value as fit quality for each simulated shower. As shown for an example event in Fig.~\ref{fig:event_fit_example}, an optimum is found using a parabolic fit in a range around the minimum $\chi^2$, and this is taken as the reconstructed \Xmax. 

The fit is now done on radio data only, in contrast to \cite{xmax:Buitink:2014} where the particle detector signals were included as well. 
This performs about equally well, while removing systematic uncertainties related to 	particle-based event reconstruction. 
A small fraction of showers (on the order of $\unit[10]{\%}$) could not be accurately reconstructed without the particle detector signals, and these are now automatically discarded. 

The primary energy of the shower is estimated from the measured radio intensity compared to the radio intensity of the best-fitting simulated shower. The radio intensity depends quadratically on the primary energy \cite{xmax:Nelles_param:2015,xmax:Zimmermann:2017}.
Therefore, the square root of the overall scale factor in the fit is used as a correction factor to the simulated energy. 
This method is now preferred, following a recent improvement in the calibration of the absolute scale of the measured radio signal \cite{xmax:Mulrey:2019}. Using the emission from the Galaxy as a calibration source, and implementing a model of the signal chain with its electronic noise contributions, allowed for a systematic uncertainty of $\unit[13]{\%}$ in amplitude. Statistical uncertainties are close to $\unit[10]{\%}$ on average.

Uncertainties on core position, energy and \Xmax are calculated from a Monte Carlo procedure on the simulated showers. 
We add to the simulated signals the noise levels measured in each antenna. For $3$ realisations of the random noise, we produce a dataset which is reconstructed as if it were a measured shower (using the same code), using the other simulated showers as reconstruction ensemble. 
We compare the real core position, \Xmax and energy, which are known for the simulations, with the reconstructions thus found. 
Their RMS errors are taken as their uncertainties, which apply to the measured shower as well as to the ensemble of simulated showers.

\section{Sample selection criteria}\label{sect:selectioncriteria}

For a composition analysis based on \Xmax and energy of a set of measured air showers, it is critical that the dataset represents the \Xmax-distribution in nature. 
Including all reconstructable measurements in the dataset typically leads to a biased sample.
Due to the irregular layout of LOFAR, it is not straightforward to establish a fiducial volume in parameter space leading to direct inclusion/exclusion criteria for measured air showers, based on shower core, arrival direction, and energy. 
Especially for moderate-sized datasets of hundreds to thousands of showers, an analysis per shower is preferred.

The main sources of bias in \Xmax arise from the thresholds of the particle detector trigger and the LOFAR radio detection. 
The particle detectors will trigger more easily for showers penetrating deeper into the atmosphere, i.e.~at high \Xmax values.
On the other hand, the LOFAR radio detection threshold, which is set at 3 LOFAR stations having a significant signal in at least half of the antennas, is easier to reach for low-\Xmax showers. As these have a larger radio footprint, they have a higher likelihood of a detection in 3 LOFAR stations.

To ensure a bias-free sample, a sufficient main criterion is that each measured shower must be able to trigger both LOFAR and the particle detector array LORA, would it have had any other \Xmax in the natural range. 
Using the Monte Carlo (simulated) ensemble for each measured shower, this is tested, as described below. 

\subsection{Cuts on reconstruction quality}
From the Monte Carlo procedure outlined in Sect.~\ref{sect:method}, we obtain uncertainties on \Xmax, energy and on the core position. These uncertainties apply to the simulated ensemble as a whole, as well as to the measured shower itself. 
By design, uncertainties calculated in this way are independent of \Xmax, and therefore no composition bias will be introduced by placing quality cuts on these uncertainties.
The uncertainty on the core position, being a basic geometric property of the shower, is used as an overall reconstruction quality indicator. 
Clearly, the uncertainties on \Xmax and core position are correlated; a cut around $\unit[7.5]{m}$ rejects most of the poorly reconstructed showers, and retains most of the good showers.

\subsection{Trigger test}
The simulated ensemble of about 30 showers provides a set of particles reaching the ground, from the Corsika simulation. 
We use a GEANT4 \cite{xmax:Geant4:2003} simulation of the particle detectors to check if these particle densities are sufficient to trigger LORA. 
Each detector will trigger if one or more particles pass through it; a coincidence of 13 out of 20 detectors is needed for a trigger. 
As this is subject to Poisson statistics, we require a trigger probability of at least $\unit[99]{\%}$.

Similarly, we require each shower in the simulated ensemble to be able to trigger 3 LOFAR stations. 
The core position from the best-fit reconstruction is assigned to all simulated showers. 
The measured noise level from the LOFAR data is taken as reference, and a signal-to-noise ratio of 6 (in intensity) is set as a condition to detect the signal for each antenna. 
From the data analysis pipeline, a station has a valid detection if at least half the antennas have a signal above this SNR.

\section{Results: test for residual bias}\label{sect:residualbias}
\begin{figure}[t]
\hspace*{-3pt}\includegraphics[width=0.51\textwidth]{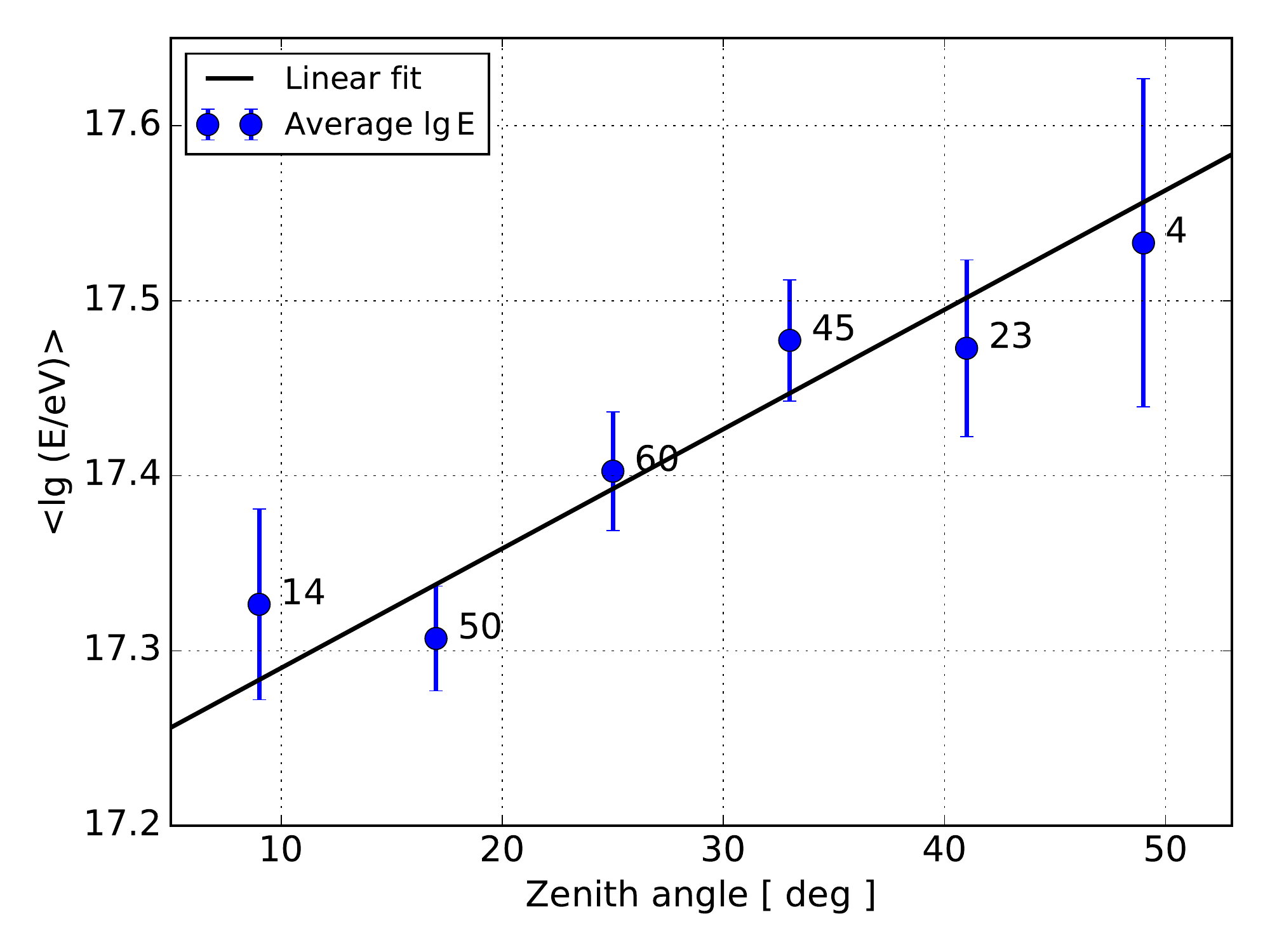}
\includegraphics[width=0.51\textwidth]{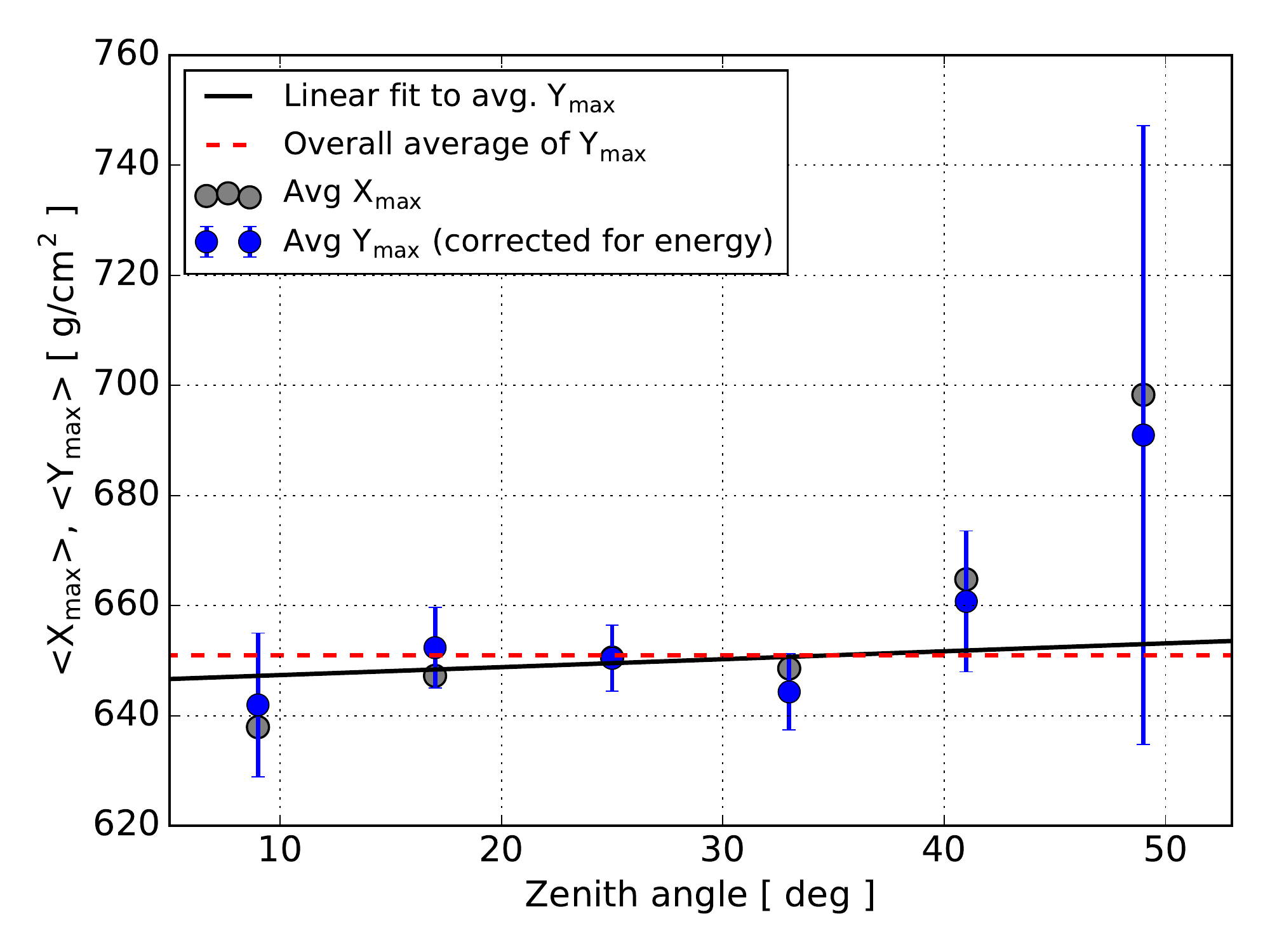}
\caption{\textbf{Left panel}: The average $\log\,E$ as a function of zenith angle, with linear fit. \textbf{Right panel}: Average energy-corrected \Xmax, defined as $Y$ in Eq.~\ref{eq:def_y}, versus zenith angle, with a constant and linear fit. }
\label{fig:lgE_xmax_vs_zenith}
\end{figure}

We have tested for bias in our data sample, after the fiducial cuts described in Sect.~\ref{sect:selectioncriteria}, by evaluating the zenith dependence of the distribution of \Xmax. 
A non-uniform value for $\langle\Xmaxmath\rangle$ over zenith angle would point at unresolved systematic effects in the events reconstruction or detection probabilities.  
However, we first need to correct for a small 
dependence of average $\lg E$ on zenith angle, as shown in Fig.~\ref{fig:lgE_xmax_vs_zenith}, left panel.
To correct for this effect we introduce a parameter $Y$ for each shower:
\begin{equation}\label{eq:def_y}
Y = \Xmaxmath + 55\,(\lg E - 17.4)\,\unit{g/cm^2},
\end{equation}
the value of 17.4 being close to the average log-energy in our sample.

The results for $Y$ as a function of zenith angle are shown in the right panel of Fig.~\ref{fig:lgE_xmax_vs_zenith}, together with a constant and a linear fit.
The slope of the linear fit is $0.14 \pm 0.44$, which is compatible with zero.
The uncertainty of the constant fit is $\unit[3.2]{g/cm^2}$, which is added to the systematic uncertainties, as a bias at this level cannot be ruled out.

\begin{figure}[h]
\begin{center}
\includegraphics[width=\textwidth]{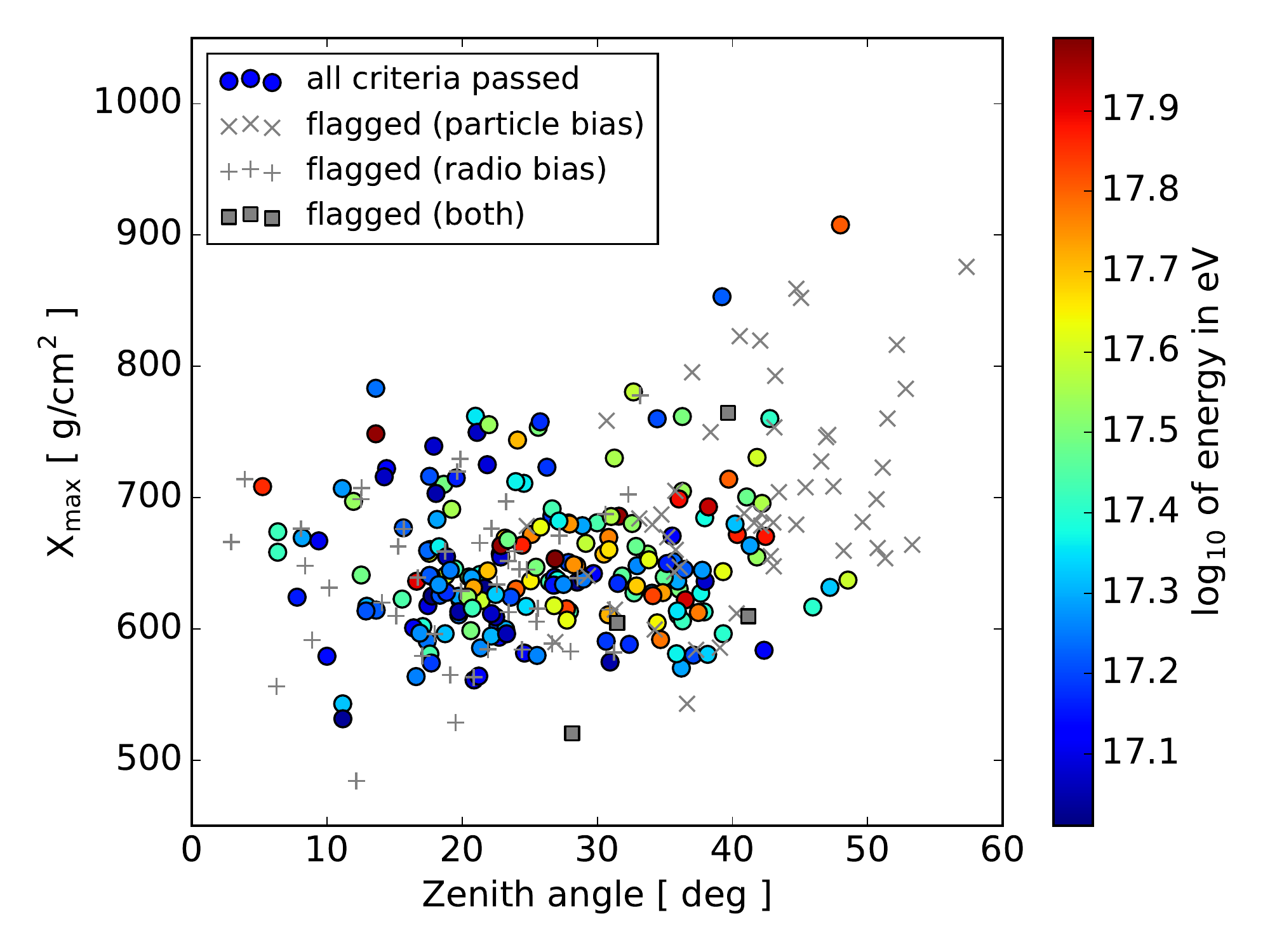}
\caption{\Xmax versus zenith angle for $298$ showers passing the reconstruction quality criterion. For those passing all criteria, the colored circles represent their energy. Showers rejected by the selection criteria for the particle trigger and the radio detection are shown as crosses and $+$ symbols.}
\label{fig:xmax_scatter_zenith}
\end{center}
\end{figure}

A complete scatter plot of \Xmax versus zenith angle is shown in Fig.~\ref{fig:xmax_scatter_zenith}, for a set of $N=298$ showers passing the reconstruction quality criterion. Of these, 196 also pass both fiducial selection criteria.

\section{Systematic uncertainties}
Systematic uncertainties in the reconstruction of \Xmax arise due to a number of factors. These are summarized in Table \ref{table:syst_uncertainties}.
Choosing a hadronic interaction model for the Corsika simulations gives a systematic uncertainty of about $\unit[5]{g/cm^2}$ \cite{xmax:Buitink:2016}; we have used QGSJetII-04 \cite{xmax:Ostapchenko:2013}. In the composition analysis, the average \Xmax for a given primary element and energy varies up to about $\unit[15]{g/cm^2}$ compared to e.g.~EPOS-LHC \cite{xmax:EPOSLHC:2013} and Sibyll-2.1 \cite{xmax:Sibyll:2009}. The latter is accounted for by doing the composition analysis separately using each model for the \Xmax distributions.

\begin{table}[b]
\caption{Systematic uncertainies in the \Xmax reconstruction}
\centering
\begin{tabular}{l l l}
\hline\hline
 & Systematic uncertainty & Additional statistical uncertainty~\\ [0.5ex] 
\hline
Choice of hadronic interaction model & $\unit[5]{g/cm^2}$ &  \\
Remaining atmospheric uncertainty & $\sim \unit[1]{g/cm^2}$ & $\sim \unit[2]{g/cm^2}$ \\
Five-layer atmosphere CORSIKA & $\unit[2]{g/cm^2}$ & $\unit[4]{g/cm^2}$ \\
Possible residual bias & $\unit[3.2]{g/cm^2}$ &  \\ 
Curve fit for $\chi^2$ optimum & $\leq \unit[1]{g/cm^2}$ &  \\
\hline
Total, added in quadrature & $\unit[7]{g/cm^2}$ & \\
\hline
\hline
\end{tabular}
\label{table:syst_uncertainties}
\end{table}

We have included local atmospheric profiles at the time of each measurement into the simulations \cite{xmax:Mitra:2019}.
This has reduced the systematic effects of density and refractive-index profiles from 4 to $\unit[11]{g/cm^2}$ \cite{xmax:Corstanje:2017} to only $\unit[2]{g/cm^2}$. 

We have tested for a possible residual bias in our \Xmax sample (see Sect.~\ref{sect:residualbias}); the amount found there has been added to the systematic uncertainties.

In the parabolic curve fit to obtain the optimum \Xmax, we have reduced a possible systematic effect by ensuring that several showers have been simulated in a close region around the reconstructed \Xmax.

In total, a systematic uncertainty on \Xmax of $\unit[7]{g/cm^2}$ is found. This is comparable to the value of $7$ to $\unit[11]{g/cm^2}$ for the Pierre Auger Observatory in this energy range \cite{xmax:Bellido:2017}.
For the energy measurement, the systematic uncertainty amounts to $\unit[14]{\%}$ \cite{xmax:Mulrey_ICRC:2019}, which is dominated by the uncertainty on the absolute scale of the radio calibration \cite{xmax:Mulrey:2019}, with two smaller contributions added in quadrature: a contribution due to the choice of hadronic interaction model, and the invisible energy in air showers, of $\unit[4]{\%}$ \cite{xmax:Aab:2019}, and a general contribution from using radiation energy from microscopic air shower simulations, of $\unit[2.6]{\%}$ \cite{xmax:Glaser:2017, xmax:Gottowik:2018}. This is an improvement of about a factor 2 over using the LORA particle detector array for the energy, which has a $\unit[27]{\%}$ systematic uncertainty \cite{xmax:Thoudam:2014, xmax:Thoudam:2016}.

\section{Summary}

We have improved the reconstruction method, by using radio data only for both the \Xmax and energy estimate. The latter is now available thanks to a better calibration of the LOFAR antennas. 
Local atmospheric profiles are included in the simulations for each shower. 
This leads to a systematic uncertainty in \Xmax of $\unit[7]{g/cm^2}$, and $\unit[14]{\%}$ in energy. 
The statistical uncertainty in the energy is reduced from about $30$ to $\unit[10]{\%}$.
The fiducial selection criteria have been improved, mainly in the evaluation of the particle trigger threshold.
We have tested for residual bias after the fiducial selection, by evaluating the zenith dependence of \Xmax in our dataset.
No significant residual bias was found, and at our level of statistics, it is bounded by $3$ to $\unit[4]{g/cm^2}$.
Thus, the current level of accuracy is sufficient for a refined composition analysis.

\begin{footnotesize}

\begin{tocless}
\bibliographystylexmax{unsrt}
\bibliographyxmax{LOFARbib}
\end{tocless}

\end{footnotesize}

\newpage


\savepagenumber
\part{The energy scale of cosmic rays detected with LOFAR}

\title{The energy scale of cosmic rays detected with LOFAR}

\author{\speaker{K. Mulrey}\!\! for the LOFAR CRKSP\\
		E-mail: \email{kmulrey@vub.ac.be}}

\abstract{The LOw-Frequency ARray (LOFAR) measures radio emission from extensive air showers.  Precise knowledge of the electric field at each antenna in the $30$-$80$\,MHz range is obtained using a newly developed, frequency-dependent calibration built on knowledge of the Galactic emission and a detailed model of the signal chain.  The energy fluence for each event is then determined, allowing for the calculation of the radiation energy of the air shower.  The radiation energy, corrected for  geometrical effects, scales quadratically with the energy contained in the electromagnetic component of the air shower.  These measurements, combined with predictions that rely only on first-principle electrodynamics, provide an energy estimate for the primary particle.  In this contribution we present the radio-based energy scale of cosmic rays detected with LOFAR, and compare it to particle-based energy measurements made using the scintillator array located at the LOFAR core.}


\begin{titlingpage}
\maketitle
\end{titlingpage}
\restorepagenumber
\addtocounter{page}{1} 

\section{Introduction}

When cosmic rays enter the Earth's atmosphere they interact, generating a cascade of secondary particles which emit coherently in the radio regime.  This emission is generated through the charge-excess effect and the transverse current induced by the magnetic field of the Earth.  The radio emission is produced primarily by the electromagnetic components of the shower, and is calculated from first principles using classical electrodynamics~\cite{escale:huege2016}.  With knowledge of the electric field at ground level, the energy fluence and radiation energy of the event can be determined.  The radiation energy, once corrected for geometrical effects, scales quadratically with the energy contained in the electromagnetic component of the air shower, thereby providing an energy estimate for the primary cosmic ray.  The measured signal is integrated over the whole air shower, and so measurements can be used to perform complete calorimetric energy reconstructions~\cite{escale:Glaser:2016qso,escale:Aab:2015vta}.

The LOw Frequency ARray (LOFAR) is well suited to study radio emission from cosmic rays because of its dense antenna spacing.  Features of the primary cosmic ray including energy, geometry, and composition are reconstructed with high precision~\cite{escale:LOFAR,escale:buitink2014,escale:schellart2013}.  The frequency range of the low-band antennas (LBAs) used is $30$-$80$\,MHz.  An in situ particle detector array, LORA, is used as a trigger for antenna readout.  Until recently, shower parameters like \Xmax were reconstructed using radio data, but the overall energy scale was set using particle data measured with LORA.  Since a new technique has been developed that provides an absolute, frequency dependent calibration for the LOFAR antennas~\cite{escale:Mulrey:2019vtz}, we move to using radio measurements to set the LOFAR energy scale \cite{escale:Corstanje:2019}.  In this contribution we present the methods by which the energies of LOFAR air showers are reconstructed, and make a comparison of the energies determined using radio and particle techniques.

\section{Method}\label{sec:method}

When LORA detects particles from an air shower, $2$\,ms of LOFAR radio data are saved.   The radio data are then cleaned for RFI and calibrated.  Finding an absolute calibration for antennas in the range $30{-}80$~MHz has been a challenge due to complications created by electronic noise and uncertainties in antenna characteristics.  To address that problem, a new technique was developed using the Galactic emission and a model of the signal chain to provide a frequency-dependent conversion factor from ADC counts to voltage at the antenna feed.  Details can be found in \cite{escale:Mulrey:2019vtz}.  The events used in this analysis have been detected by at least three LOFAR stations, and passed quality cuts on polarization and geometrical reconstruction.

The combination of emission mechanisms in the air showers leads to a complicated radio footprint at ground level.  The emission mechanisms are well understood, and the pattern can be reproduced with particle-level, first principle simulations like CORSIKA~\cite{escale:corsika} and the radio plug-in CoREAS~\cite{escale:Huege:2013vt}.  In this section two methods of determining cosmic-ray energy based on radio measurements are described.  The first is referred to as the direct scaling method, and is what has been used in past LOFAR analyses.  The second is a fluence-based method, which follows the approaches described in~\cite{escale:Glaser:2016qso} and~\cite{escale:Aab:2015vta}.  Both methods are based on CoREAS simulations.

\subsection{Direct scaling method}

For each LOFAR event, a set of simulations is generated covering a range of $X_{\mathrm{max}}$ values for both proton and iron primaries (CORSIKA v-7.6300~\cite{escale:corsika}, FLUKA~\cite{escale:fluka}, QGSJetII-04~\cite{escale:qgsjet}).  The arrival direction used in the simulation is based on timing reconstruction of the event, and the energy used in the simulation is based on the results of a two dimensional LDF fit to the radio data~\cite{escale:nelles2015}.  Each simulation includes a realistic atmosphere, specific to the time and location of the event~\cite{escale:Mitra:2019}.  Antennas are simulated in a star shape pattern in the antenna plane, and the LOFAR LBA antenna model is applied to the simulated electric field, yielding the voltage at the antenna.  The integrated power within a $55$\,ns window is calculated. This is interpolated into a two dimensional map.  Full details of the method are found in~\cite{escale:buitink2014}.  This map is then fit to LOFAR data using a minimization procedure with free parameters for the core position of the shower and a scale factor for the energy, as

\begin{equation}\label{eq:chi2}
    \chi^2=\sum_{\mathrm{antennas}}\bigg( \frac{P_{\mathrm{ant}}-f_r^2P_{\mathrm{sim}}(x_{\mathrm{ant}}-x_0,y_{\mathrm{ant}}-y_0)}{\sigma_{\mathrm{ant}}}  \bigg)^{\!\!2}
\end{equation}
where $P_{\mathrm{ant}}$ is the measured integrated power,  $P_{\mathrm{sim}}$ is the simulated integrated power at ($x_{\mathrm{ant}},y_{\mathrm{ant}}$), $\sigma_{\mathrm{ant}}$ is the noise level in the antenna, ($x_0,y_0$) is the core position, and $f_r$ is the energy scaling factor.   The simulation with the lowest $\chi^2$  value is taken as the ``truth.'' The cosmic-ray energy, $E_{\mathrm{dir. scal.}}$,  is determined by multiplying the simulated energy by the scale factor.

\subsection{Fluence-based method}

The fluence-based method uses the best-fit simulation as determined by the direct scaling method.  With the CoREAS simulation closest to the true event, we follow the procedure outlined in~\cite{escale:Glaser:2016qso} to determine the radiation energy.
The energy fluence at each antenna is calculated using
\begin{equation}
    f(\vec{r})=\epsilon_0 c \Delta t \sum_i E^2(\vec{r},t_i)
\end{equation}
where $c$ is the speed of light in vacuum, $\epsilon_0$ is the vacuum permittivity, and $\Delta t$ is the sampling interval of the time and position dependent electric field $E(\vec{r},t)$.
The radiation energy is then calculated as
\begin{equation}
    E_{\mathrm{rad}}=\int^{2\pi}_0\!\!\!\mathrm{d}\phi \int^{\infty}_0 \!\!\!\mathrm{d}r\, r f(r,\phi)
\end{equation}
where we make use of the radial symmetry of the geomagnetic and charge excess distributions, and use the fluence where $\phi$ is along the $\vec{v}\times\vec{v}\times\vec{B}$ axis in the shower plane.   This results in an overestimate of the radiation energy by $3.36\%~\cite{escale:Gottowik:2017wio}$, which is corrected for.  The strength of the radio emission depends on the local magnetic field, which affects the geomagnetic component of the emission.  The radiation energy is corrected by

\begin{equation}\label{eq:srd}
    S_{RD}=\frac{E_{\mathrm{rad}}}{\big({a'}^2+(1-{a'}^2)\big)\,\mathrm{sin}^2\alpha\, \big(\frac{B_{\mathrm{Earth}}}{0.243 \mathrm{G}}\big)^{\!\!1.8}}
\end{equation}
where $a' = a/(B_{\mathrm{Earth}}/0.243\mathrm{G})^{0.9}$, $a$ is a parametrization of the charge-excess fraction based on the simulated \Xmax, and $\alpha$ is the angle between $\vec{v}$ and $\vec{B}$.  $B_{\mathrm{Earth}}=0.496$~G is the local magnetic field at LOFAR and $0.243$~G is the magnetic field strength at the Auger site. Details of Eq.\,\ref{eq:srd} and the parametrization of the charge-excess fraction are found in~\cite{escale:Glaser:2016qso}.
The radiation energy is increased by 11\% to correct for the electron  multiple scattering length factor in the underlying EGS4 simulation, following~\cite{escale:Gottowik:2017wio}.  To compare to measured data, the radiation energy is scaled by $f_r^2$ from Eq.\,\ref{eq:chi2}.

We next find conversions between the radiation energy and the electromagnetic (energy deposited in the atmosphere by electromagnetic components of the shower) and total cosmic ray energies based on CoREAS simulations.  Simulations for existing LOFAR analyses are used, consisting of 4900 proton and 4900 iron showers, ranging from $10^{16.5}{-}10^{18}$ eV, with zenith angles between $0^{\circ}{-}\,50^{\circ}$, and azimuth angles between $0^{\circ}{-}\,360^{\circ}$.  We find the radiation energy for each as described above and relate them to the electromagnetic energy and primary cosmic ray energy by fitting the power law equation 
\begin{equation}\label{eq:srd_to_e}
    S_{RD} = A \times 10^7 \mathrm{eV} \left(\frac{E}{10^{18} \mathrm{eV}}\right)^{\!\!B}.
\end{equation}

\begin{figure}[h]
\centering
\includegraphics[scale=0.4]{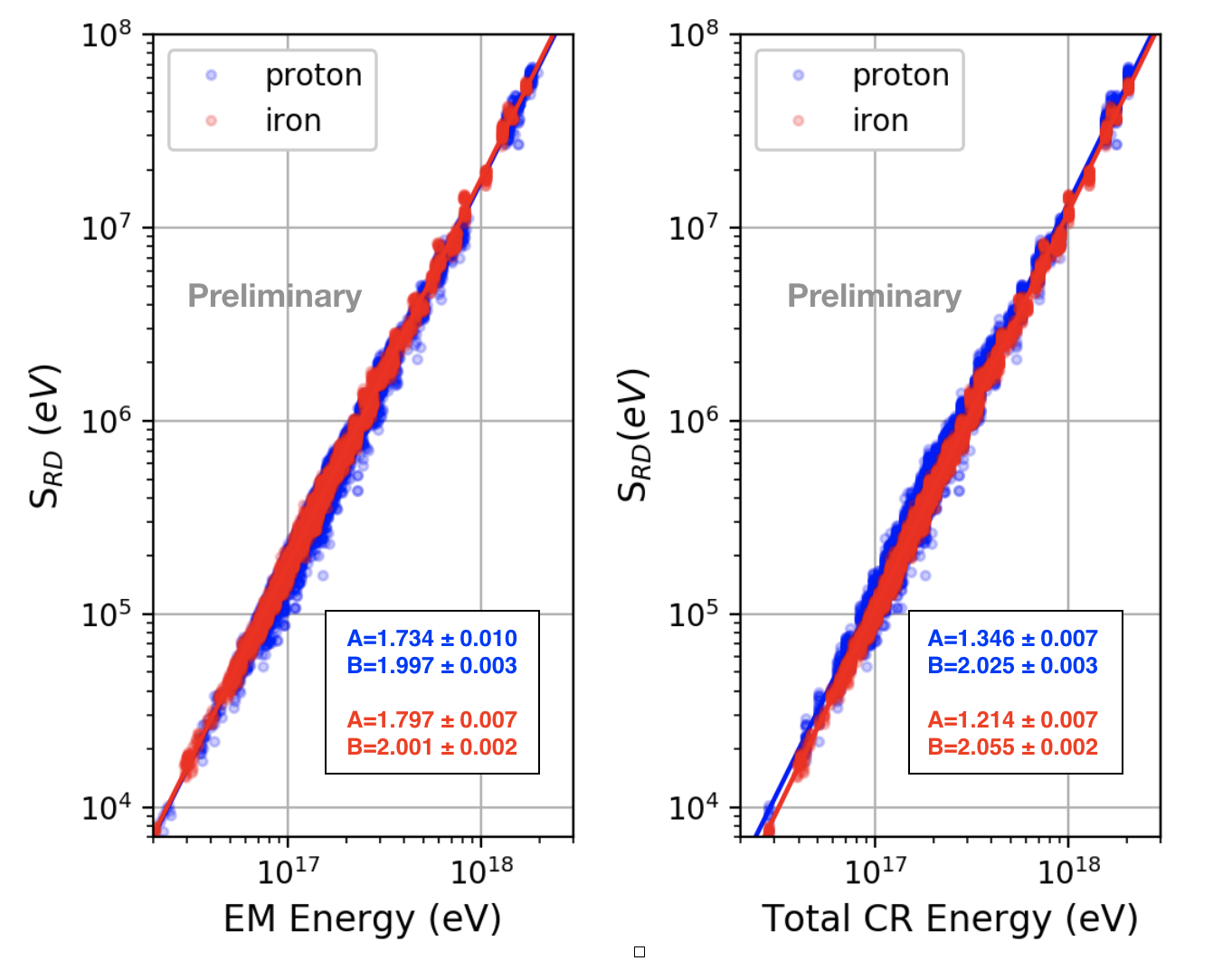}
\caption{Radiation energy as a function of electromagnetic energy \textbf{(left)} and total air shower energy \textbf{(right)}.  Proton events are shown in red, and iron in blue.  Parameters A and B are best fit values for Equation \ref{eq:srd_to_e}, and are detailed in Table~\ref{tab:fitval}.}
\label{fig:em_total}
\end{figure}

The results of this fit are shown in Fig.\,\ref{fig:em_total}.  The radiation energy as a function of the electromagnetic energy is shown in the left panel, and as a function of total cosmic-ray energy in the right panel.  The radiation energy is expected to scale quadratically with the energy contained in the electromagnetic components of the shower, and indeed that is seen.  In the case of total energy, the difference between primary species is more evident.  This is due to the fact that the total cosmic-ray energy includes the invisible energy from particles that do not release all their energy into the atmosphere, which is not measurable by radio antennas~\cite{escale:linsley1983}.

The values for fit parameters A and B found for electromagnetic energy are similar to those found in~\cite{escale:Glaser:2016qso}, and detailed in Table~\ref{tab:fitval}.  Differences in the parameters may be due to the fact that a second order correction due to atmospheric density used in~\cite{escale:Glaser:2016qso} is not included here.  Also, the simulation sets in each case are not identical, including different energy ranges, zenith angles, and locations, although most differences are accounted for by the corrections made in Eq.\,\ref{eq:srd}.

\begin{table}
\caption{Best-fit parameters of Eq.\,\ref{eq:srd_to_e} for simulated proton and iron showers, which yield an energy estimation for the electromagnetic and total energy contained in the shower.}
\begin{center}\label{tab:fitval}
 \begin{tabular}{||c c c||} 
 \hline
 &A & B  \\ [0.5ex] 
 \hline\hline
 Proton EM energy  & 1.734 $\pm$ 0.010 & 1.997 $\pm$ 0.003 \\ 
 \hline
 Iron EM energy  & 1.797 $\pm$ 0.007 & 2.001 $\pm$ 0.002  \\
 \hline
 mixed EM energy from \cite{escale:Glaser:2016qso} & 1.629$\pm$ 0.003 & 1.98 $\pm$ 0.001  \\
 \hline
 Proton total energy & 1.346 $\pm$ 0.007 & 2.025 $\pm$ 0.003 \\
 \hline
 Iron total energy & 1.214 $\pm$ 0.007 & 2.055 $\pm$ 0.002  \\ [1ex] 
 \hline
\end{tabular}
\end{center}
\end{table}

\section{LOFAR energy scale}
In this section we study the energy scale of LOFAR events based on radio and particle data and make a comparison of the two. 

\subsection{Radio-based energy scale}

\begin{figure}[h]
\centering
\includegraphics[scale=0.42]{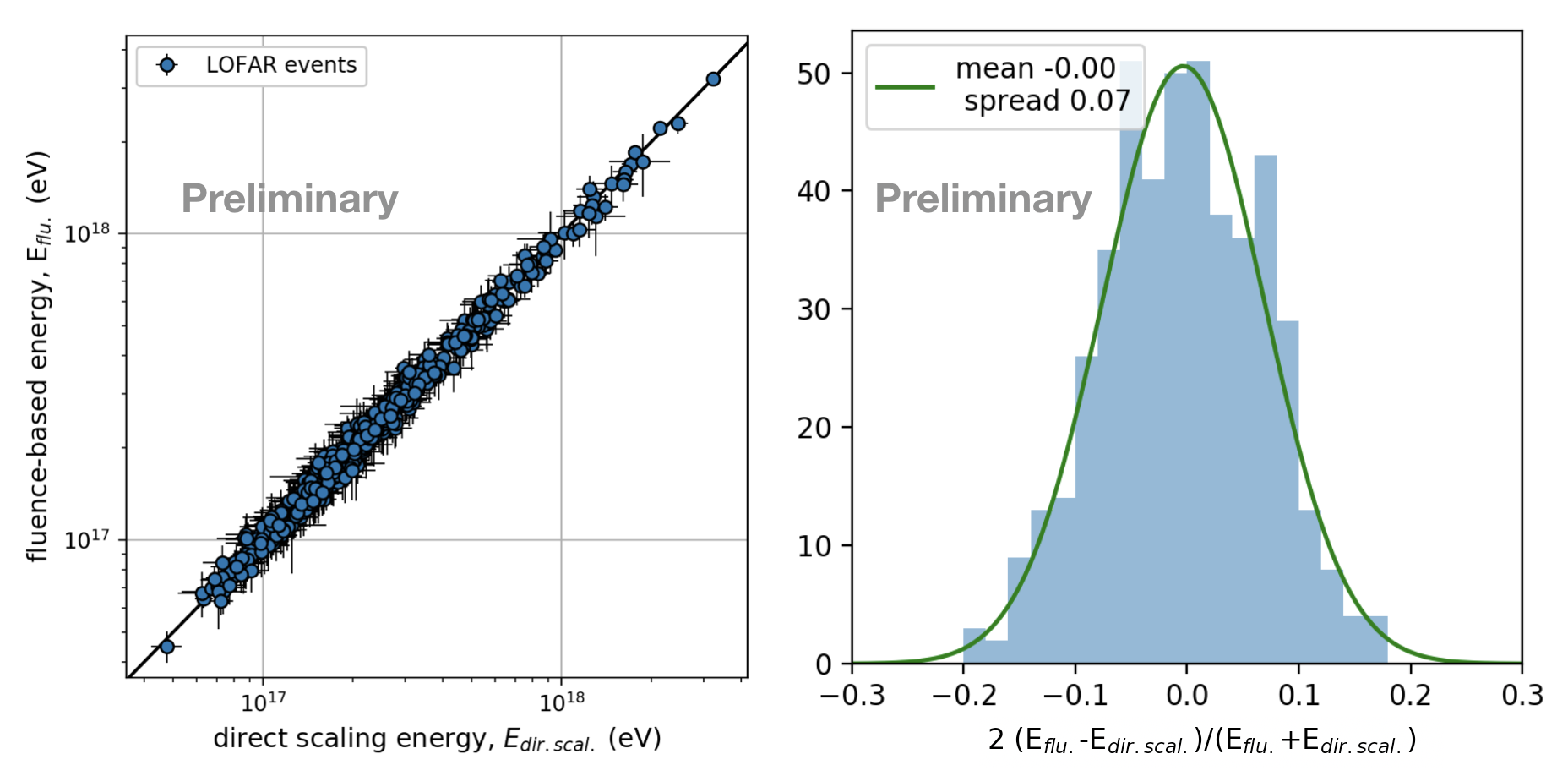}
\caption{\textbf{Left:} Comparison of the energy found using the direct scaling method and the fluence-based method.  Error bars indicate systematic uncertainties.  \textbf{Right:}  Relative differences between methods.}
\label{fig:radio_compare}
\end{figure}

In order to calculate the total cosmic-ray energy on an event-to-event basis using Eq.\,\ref{eq:srd_to_e}, the primary composition has to be taken into account, as there is a difference between proton and iron fit parameters for total energy fits.  We use total energy, and not electromagnetic energy, because we will compare the radio-based energy measurements with the particle-based measurements which yield the total cosmic-ray energy.  For each event, we choose the composition of the best fit simulation, as described in Section~\ref{sec:method} and~\cite{escale:buitink2014}.  We first check that fluence-based method of determining energy is consistent with the direct scaling method.  This is shown in Fig.\,\ref{fig:radio_compare}.  The left panel shows a scatter plot of the cosmic-ray energy for each LOFAR event found using the direct scaling method ($E_{\mathrm{dir. scal.}}$) vs the fluence-based method ($E_{\mathrm{flu.}}$).  The right panel shows the relative differences between the two methods, where there is no offset.

\subsubsection{Uncertainties on radio-based energy scale}

The antenna calibration is the largest contribution to the systematic uncertainties, at 13\%~\cite{escale:Mulrey:2019vtz}.  Uncertainties in the fit procedure have been studied in \cite{escale:buitink2014}, where it was shown that the method resulted in no systematic offsets in the reconstructed core position or energy.  While the electromagnetic energy of the shower has little dependence on hadronic interaction models, some uncertainty is introduced when we use the total shower energy.  From~\cite{escale:Aab:2019cwj}, we see that the differences in invisible (non-calorimetric) energy due to hadronic interaction models are less than 4\% for energies above $10^{17}$~eV, which we take as a systematic uncertainty on choice of model.    In general, the uncertainties introduced using radiation energy from particle-level simulations translate into 2.6\% in cosmic-ray energy~\cite{escale:Glaser:2016qso,escale:Gottowik:2017wio}.  The systematic uncertainties sum in quadrature to 14\%.

The statistical uncertainties are derived from a Monte Carlo study.  For each event, a set of simulations is produced with the same energy and geometry.  One simulation is taken to be truth, remaining simulations
are fitted to it in order to reconstruct the energy and core position. This method is repeated for all the showers in the ensemble.  Each reconstruction results in a scale factor $f_r$ from Eq.\,\ref{eq:chi2}.  The standard deviation of the set of scale factors~$f_r$ sets the statistical uncertainty of the event energy. 

\subsection{Particle-based energy scale}

We compare the energy scale now established with the fluence-based method with the energy derived from particle measurements.  The particle energy scale is also derived using a simulation-based approach.  The best fitting CORSIKA simulation as determined by the $X_{\mathrm{max}}$ analysis is used.  The particle distribution on the ground is binned into distance from shower core, and a GEANT4~\cite{escale:geant4} simulation of the LORA detectors converts particles that would produce a signal to energy deposited in the scintillators~\cite{escale:buitink2014,escale:thoudam2014}.  One challenge with the LORA particle data is that shower cores outside the superterp are difficult to constrain based on particle data alone.  For this reason, the radio-based core location is used.  Then, a particle $\chi^2$ is fit, as
\begin{equation}
    \chi^2=\sum_{\substack{\mathrm{particle}\\ \mathrm{detectors}}}\left( \frac{d_{\mathrm{det}}-f_p d_{\mathrm{sim}}}{\sigma_{\mathrm{det}}} \right)^{\!\!2}
\end{equation}
where, $d_{\mathrm{det}}$ is the
deposited energy measured by a LORA detector with noise $\sigma_{\mathrm{det}}$, and $d_{\mathrm{sim}}$ is the GEANT4 simulated deposit.  The particle energy, $E_{\mathrm{part.}}$, is then the simulated energy multiplied by the particle scale factor $f_p$.

\subsubsection{Uncertainties on particle-based energy scale}

Systematic uncertainties on the particle-based energy scale are estimated following \cite{escale:thoudam2014}.  Uncertainties in the deposited energy are on the order of 30\%.  There are also uncertainties in the losses in the signal chain (cable losses, reflections) that contribute 15\%.  These two contributions yield a systematic uncertainty of 33\% on energy.  Statistical uncertainties are handled in the same manner as for radio data. The set of simulations produced for the fitting procedure is used to reconstruct a mock data set, resulting in  a set of $f_p$ scale factors, the standard deviation of which is used to set the statistical uncertainty for the event energy.

\subsection{Comparison of radio and particle energy scales}
\begin{figure}[h]
\centering
\includegraphics[scale=0.42]{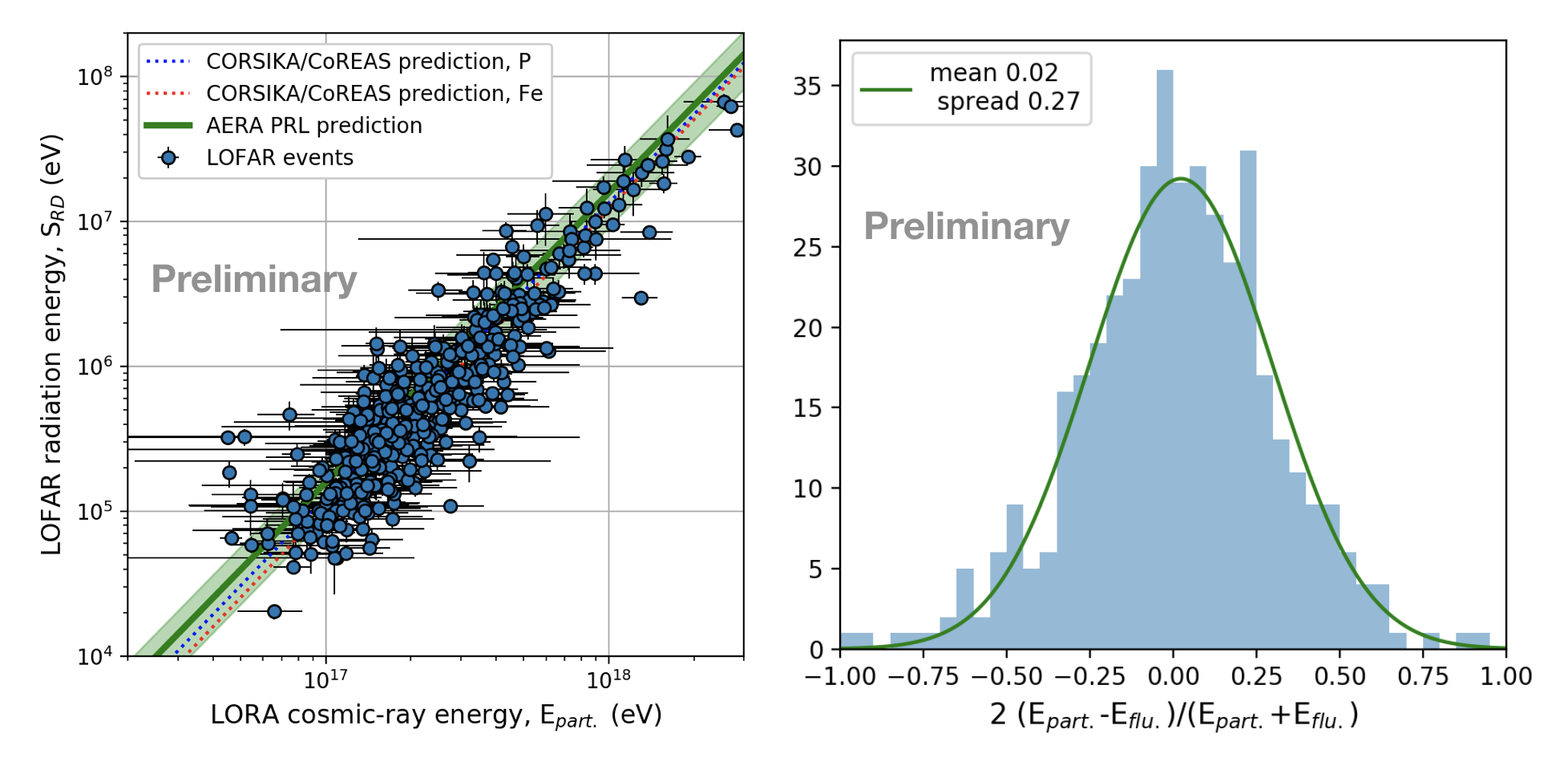}
\caption{\textbf{Left:}  Radiation energy as a function of cosmic-ray energy determined by LORA measurements. Also shown are predictions from CORSIKA/CoREAS with fit parameters described above, and expectations based on AERA results.  \textbf{Right:}~Relative difference between cosmic-ray energy determined with radio and particle data.}
\label{fig:radio_particle}
\end{figure}

The comparison between radio-based energy and particle-based energy is shown in Fig.\,\ref{fig:radio_particle}.  In the left panel, the LOFAR radiation energy ($S_\mathrm{RD}$) is shown as a function of the cosmic-ray energy derived from the LORA scintillators ($E_{\mathrm{part.}}$).  The error bars indicate statistical uncertainties.  The predictions for radiation energy based on the total cosmic-ray energy from the fit parameters for proton and iron showers (from Eq.\,\ref{eq:srd_to_e}) are shown as blue and red dashed lines.  The radiation energy prediction from the AERA experiment is shown in green~\cite{escale:Aab:2015vta}.  The right panel shows the relative differences in total cosmic-ray energy reconstructed with radio fluence ($E_{\mathrm{flu.}}$) and particle ($E_{\mathrm{part.}}$) methods.  

\section{Summary}

It has been shown that energy reconstruction using the radiation energy provides a universal, calorimetric energy measurement~\cite{escale:Glaser:2016qso,escale:Aab:2015vta}.  We have now established that the LOFAR energy scale measured this way is consistent with previous radio-based measurements, and also with the particle-based LORA energy scale.  Understanding the energy scale has important consequences for the interpretation of other air shower properties measured with LOFAR, such as \Xmax. Going forward, this technique will also allow measurements from different experiments to be directly compared.  

\begin{footnotesize}

\begin{tocless}
\bibliographyescale{LOFARbib}
\bibliographystyleescale{unsrt}
\end{tocless}

\end{footnotesize}

\newpage


\savepagenumber
\part{Reconstructing air showers with LOFAR using event specific GDAS atmospheres}

\title{Reconstructing air showers with LOFAR using event specific GDAS atmospheres}

\author{\speaker{P. Mitra}\!\! for the LOFAR CRKSP\\
		E-mail: \email{pmitra@vub.be}}

\abstract{Estimating the depth of shower maximum \Xmax with high precision 
is of great interest for the study of primary particle composition. One of the systematic
uncertainties in reconstructing \Xmax from the radio emission of air 
showers is the limited knowledge of the atmospheric parameters like humidity,  
pressure, temperature and the index-of refraction. Using the Global Data Assimilation System (GDAS), a global atmospheric model, 
we have implemented time-dependent realistic atmospheric profiles in the air shower simulation codes CORSIKA and the radio plug-in CoREAS.
This program is now available within CORSIKA and flexible to be adapted for different air shower experiments.
We have analyzed the LOFAR cosmic ray data with dedicated simulations for each detected air shower with event specific GDAS atmospheres 
and investigated the effects of pressure and humidity on the reconstructed \Xmax.
This study shows that for bulk of the events, where the ground pressure is close to US standard atmosphere values,
there is a small systematic shift in \Xmax that is less than 2 $\mathrm{g/cm^{2}}$ and for very low 
pressure values the shift is up to 15 $\mathrm{g/cm^{2}}$.}

\begin{titlingpage}
\maketitle
\end{titlingpage}
\restorepagenumber
\addtocounter{page}{1} 

\section{Introduction}
The radio detection of cosmic rays is a promising technique providing deeper insight in the development of extensive air showers. 
Estimating the depth of shower maximum \Xmax with improved accuracy is 
of great interest for the study of the primary particle composition \cite{gdas:nature}.
Thus, the knowledge of atmospheric variables like temperature, humidity and pressure are important
for the estimation of \Xmax.
The highest precision of the measurement of \Xmax with the radio technique
has been currently achieved with the LOFAR radio telescope, situated in the north of the 
Netherlands. The dense core of LOFAR  consists of
288 low-band antennas within a diameter of 320 meters,
recording cosmic ray events in the 30-80~MHz band.
This provides the 
opportunity to investigate  the radio footprint, i.e.the
lateral intensity distribution in close detail and enables us
to infer \Xmax to a precision of 20 $\mathrm{g/cm ^{2}}$ 
which is comparable to the precision achieved with the fluorescence detectors \cite{gdas:xmax}.
The precision of the reconstruction is sensitive to the choice of an atmospheric model included in the Monte Carlo air shower
simulation codes.
The measured lateral intensity distributions are compared to 
CoREAS \cite{gdas:tim}, a simulation package for the radio 
emission from the individual particles in the cascade simulated
with CORSIKA .There are several options for including parameterized atmospheres for different observatories \cite{gdas:corsikamanual}. The refractive index of air 
also plays a crucial role in the radio emission of cosmic rays. It determines the propagation
velocity of the radio signal at different altitudes and causes highly time-compressed signals on ground for observers
located at the corresponding Cherenkov angle. The angle for Cherenkov emission is
sensitive to the refractive index $n$. Thus variations in the refractive index lead
to changes in the radio intensity footprint on the ground \cite{gdas:Arthurpaper}. 
Therefore, a realistic description of the atmospheric density and refractive index profiles needs to be incorporated in CoREAS.
In LOFAR analysis, dedicated simulation sets for each detected shower are used. This requires that atmospheric
profiles be included in the simulation which closely matches the conditions during the time of
actual measurements for an accurate estimation of \Xmax.

We report the results of a study on the effects of atmospheric parameters like pressure and humidity on the
reconstructed \Xmax analyzing LOFAR data,by including event-specific atmospheric conditions using
GDAS\footnote{%
Global Data Assimilation System, operated by the US National Ocanic and Atmospheric Administration\\ (\,https://www.ncdc.noaa.gov/data-access/model-data/model-datasets/global-data-assimilation-system-gdas\,).
}, %
a global meteorological model. For this purpose we have developed  a program that extracts GDAS atmospheric parameters which is interfaced with CORSIKA. This program
is available for public use since the release of CORSIKA version 7.6300. It is flexible and can be adapted by the users to obtain parameterized
atmospheric profiles for user-specified times and locations. 
\section{Atmopsheric profiles from GDAS}
The  Global Data  Assimilation  System (GDAS)  
is a  database of atmospheric data used
for weather forecasting. 
It provides data, every three hours, for several atmospheric state variables packed in $1^{\circ}\times 1^{\circ}$ latitude longitude grid.
For air shower analysis important parameters are temperature, pressure,
relative humidity, air density, and atmospheric depth at several altitude levels. The first three
quantities and the altitude are directly available in the GDAS data  and are used to calculate
air density and atmospheric depth.
Next, the refractivity $N$ defined as $N=\left(n-1\right) 10^{6}$ can be expressed as
as a function of humidity, pressure and temperature \cite{gdas:Arthurpaper}
\begin{equation}
 N = \SI{77.6890}{\kelvin\per\hecto\pascal}\frac{p_{d}}{T} + \SI{71.2952}{\kelvin\per\hecto\pascal}\frac{p_{w}}{T} + \SI{375463}{\square\kelvin\per\hecto\pascal}\frac{p_w}{T^{2}}
 \label{ri}
\end{equation}
with $p_{w}$, $p_{d}$ and $T$ being the partial water
vapor pressure, partial dry air pressure, and temperature, respectively.
The effect of humidity is important for our study as it tends to increase the refractivity in comparison to that of dry air
in the radio frequency regimes. We studied several atmospheric profiles at LOFAR. Fig.\,\ref{atmoprofile} (left)  shows the mean profile for the relative difference in refractivity $\Delta{N}_{\mathrm{relative}}$ between GDAS and the US standard atmosphere as a function of altitude 
for over 3 years consisting of 100 atmospheric profiles that correspond to the time of recorded cosmic ray events at LOFAR. 
It is defined as $\Delta{N}_{\mathrm{relative}} = (N_{\mathrm{GDAS}}-N_{\mathrm{US}})/N_{\mathrm{US}}$, $N_{\mathrm{GDAS}}$ is calculated from
Eq.\,\ref{ri} using GDAS atmospheres at LOFAR and $N_{\mathrm{US}}$ is obtained from a linear scaling of the density profile using US standard atmosphere with respect to a 
fixed ground level refractivity. 
The absolute value of the mean $\Delta{N}_{\mathrm{relative}}$
is around $10\%$ near ground and around  $3\textendash 8 \%$ between 3 and 10 km of altitude, the region important for shower development.
To account for the propagation effects of the radio signal, it is important to calculate the effective integrated refractive index between the point of emission and the observer.
The typical values for relative effective refractive indices are around  7-10\% from the altitude range 3-10 km. 
Fig.\,\ref{atmoprofile}({right) 
shows the  difference in atmospheric depth profile between the US standard atmosphere and the GDAS atmospheres at LOFAR
for 8 profiles over the years 2011-2016. The GDAS atmospheres 
vary significantly from the US atmosphere. Atmospheric profiles with similar
ground atmospheric depth can evolve differently higher in the atmosphere. This is important for calculating
the correct geometric distance to the shower maximum. 
\begin{figure}[h!]
\includegraphics[width=0.5\linewidth]{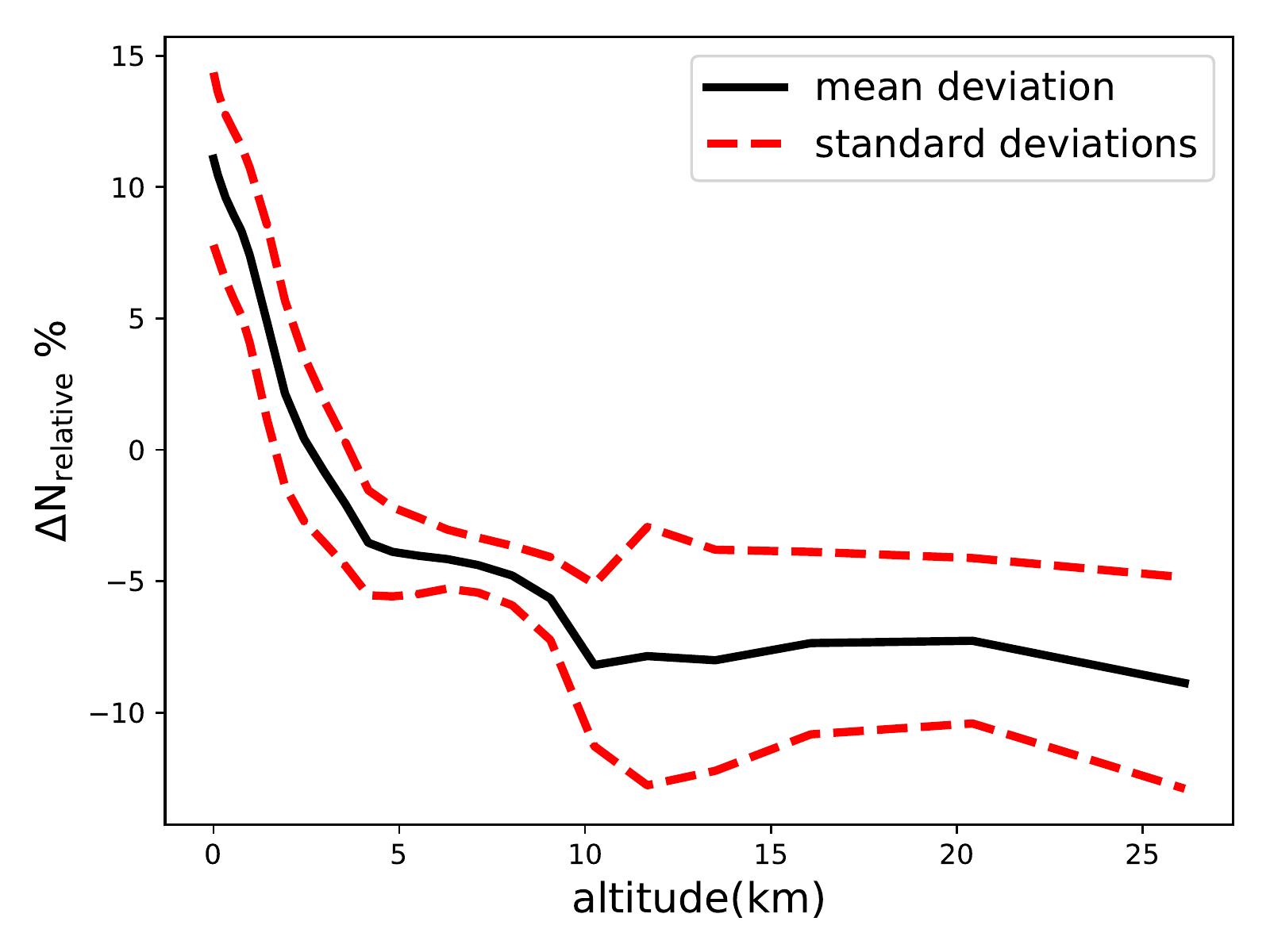}
\includegraphics[width=0.5\linewidth]{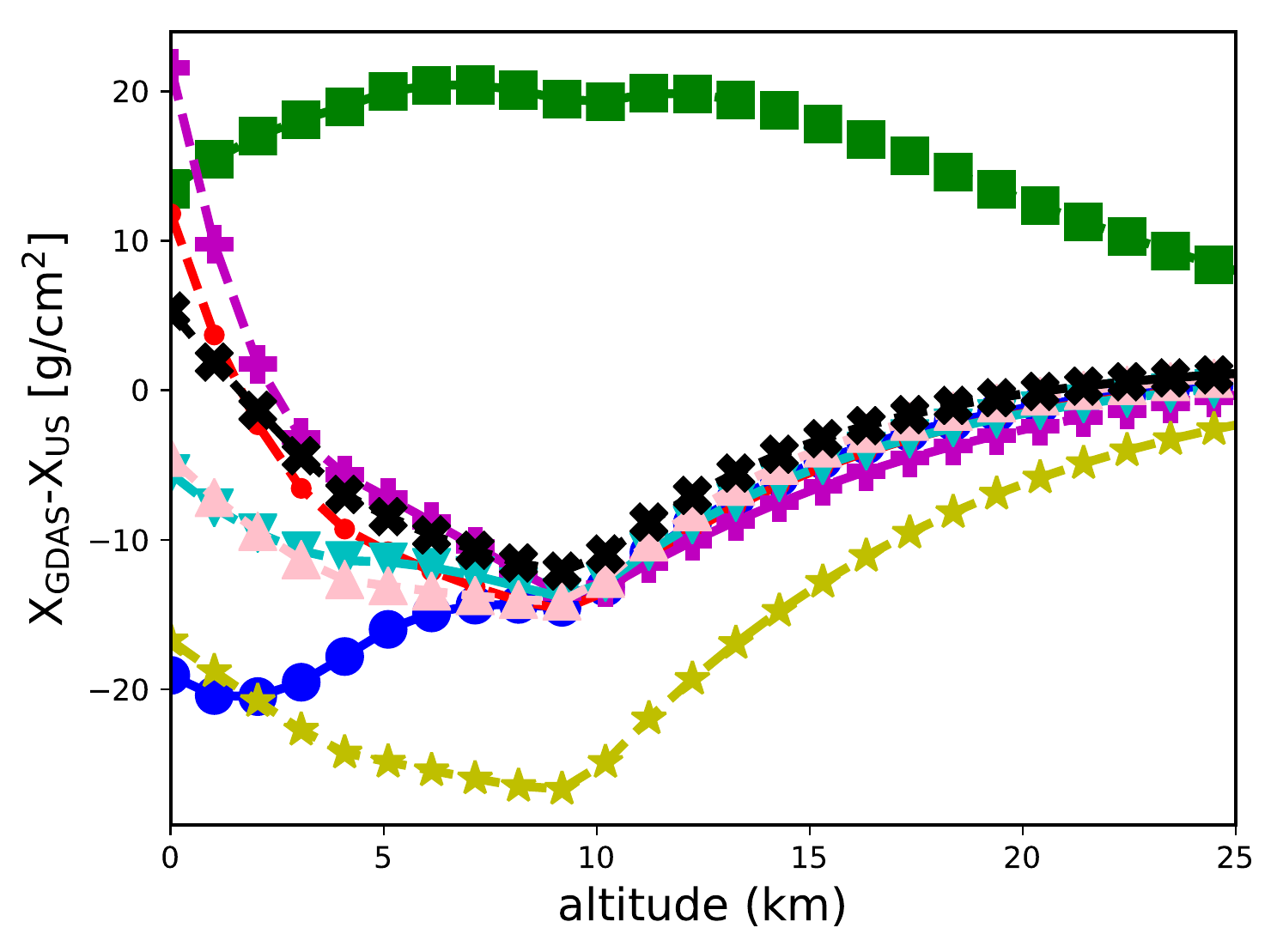}
\caption{\textbf{Left}: Mean relative refractivity as a function 
  of altitude for 100 atmospheric profiles at LOFAR spanning over the years 2011 to
   2014. The black solid line denotes the mean profile and the red dashed lines show the standard deviations.
\textbf{Right}: profiles for the difference in atmospheric depth  between 
  US standard atmosphere and GDAS atmospheres at LOFAR as a function of altitude between the years 2011-2016.}
\label{atmoprofile}
\end{figure}

\section{Implementation of GDAS in CORSIKA/CoREAS}
We have developed a program called \textquoteleft{gdastool}\textquoteright\
 that extracts the required GDAS atmospheric parameters
given the coordinates of an observatory and UTC timestamp
of the event. It then returns the parameterized density profile
fitted with the layered atmospheric model used in CORSIKA with proper boundary conditions and additionally a
table for refractive index including humidity effects as a function of height in \SI{1}{\meter} step size. The option for
GDAS parameterized atmosphere can be invoked by a new keyword \textquoteleft{ATMF}\textquoteright\ in the CORSIKA input file.
This feature is available as of CORSIKA version 76300.

\section{Results and analysis}
We have analyzed LOFAR events simulated with new GDAS atmospheres
and compared  the reconstructed \Xmax\ \cite{gdas:xmax} between various simulation sets

\begin{description}
 \item[\textbf{Set A} \textemdash] the showers were simulated with CORSIKA v-7.6300 and GDAS atmosphere including humidity effects.
 \item[\textbf{Set B} \textemdash] the showers were simulated with CORSIKA v-7.4385 and US standard atmosphere. 
 \item[\textbf{Set C} \textemdash] this set is identical to Set B, but with an additional correction factor to correct for 
 realistic atmospheres posterior to the reconstruction. This is done by reconstructing the geometrical distance to the 
 shower maximum using CoREAS simulations with US standard atmospheres, and then calculating the corresponding value 
 for  using a realistic GDAS atmosphere. This correction compensates for changes in the total mass overburden, but 
 not \Xmax for effects related to changes in the index of refraction. This method was used in previous LOFAR analysis \cite{gdas:xmax}.
\end{description} 
\begin{figure}[h!]
\begin{center}
\includegraphics[width=0.7\textwidth]{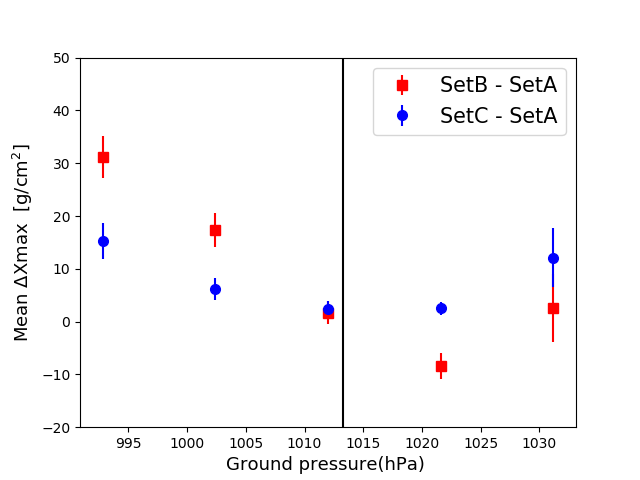}
\caption{Difference in mean \Xmax as a function of ground pressure. The total sample contains
123 air showers recorded at LOFAR. The black line denotes the U.S standard atmospheric pressure.  }
\label{Pplot}
\end{center}
\end{figure}
In Fig.\,\ref{Pplot} the difference in mean reconstructed \Xmax between the various simulation sets mentioned above is plotted against
ground pressure obtained from GDAS. Both the blue circles and red squares converge to zero where GDAS pressure approaches the US standard pressure
at 1013 hPa. 
The red squares have large $\Delta\Xmaxmath$ in general. This is expected from the fact that there is no atmospheric correction involved in Set-B. 
The blue circles  however show a significant deviation both at low and high  pressure values. This suggests that the linear first order correction added
to the standard US atmosphere implemented in Set-C is not sufficient. Full GDAS-based atmospheric profiles become indispensable when the air pressure 
at the ground deviates by more than ${\sim}\,10\,$hPa from the US standard atmosphere value.

\subsection{Effects of humidity}

In the radio frequency regime, the humidity results in higher values of
refractive index. For this study, two sets of simulations were produced.
In one set the shower was simulated with a GDAS atmosphere with extreme humid weather conditions and in the other
with a GDAS atmosphere with zero humidity. The same atmospheric
parameters are used in both cases to ensure that the particles evolve in a similar way in the atmosphere and produce same
shower maximum. In this way the inclusion of humidity only influences the simulated radio pulses.
The difference in the refractive index  contributes to difference in propagation effects on the pulse arrival time
and power. The lateral distribution of the energy fluence, the time-integrated power per unit area, for different observer
positions is also studied for different frequency bands for these two cases, as shown in Fig.\,\ref{cheren}. In the low frequency band of $30$-$80$\,MHz
relevant for LOFAR, the difference in the fluence between the two sets is small, up to the order of 4$\%$. In the high frequency band of $50$-$350$\,MHz, corresponding to the SKA-low band, the
values are larger, being around 10$\%$. In the higher frequency band the Cherenkov-like effects
become stronger and the signal is compressed along the Cherenkov ring \cite{gdas:nelles2015}. 
The opening angle is strongly dependent on the index of refraction. 
This explains the higher difference in power in Fig.\,\ref{cheren}.\\
\begin{figure}[h!]
\includegraphics[width=0.515\linewidth]{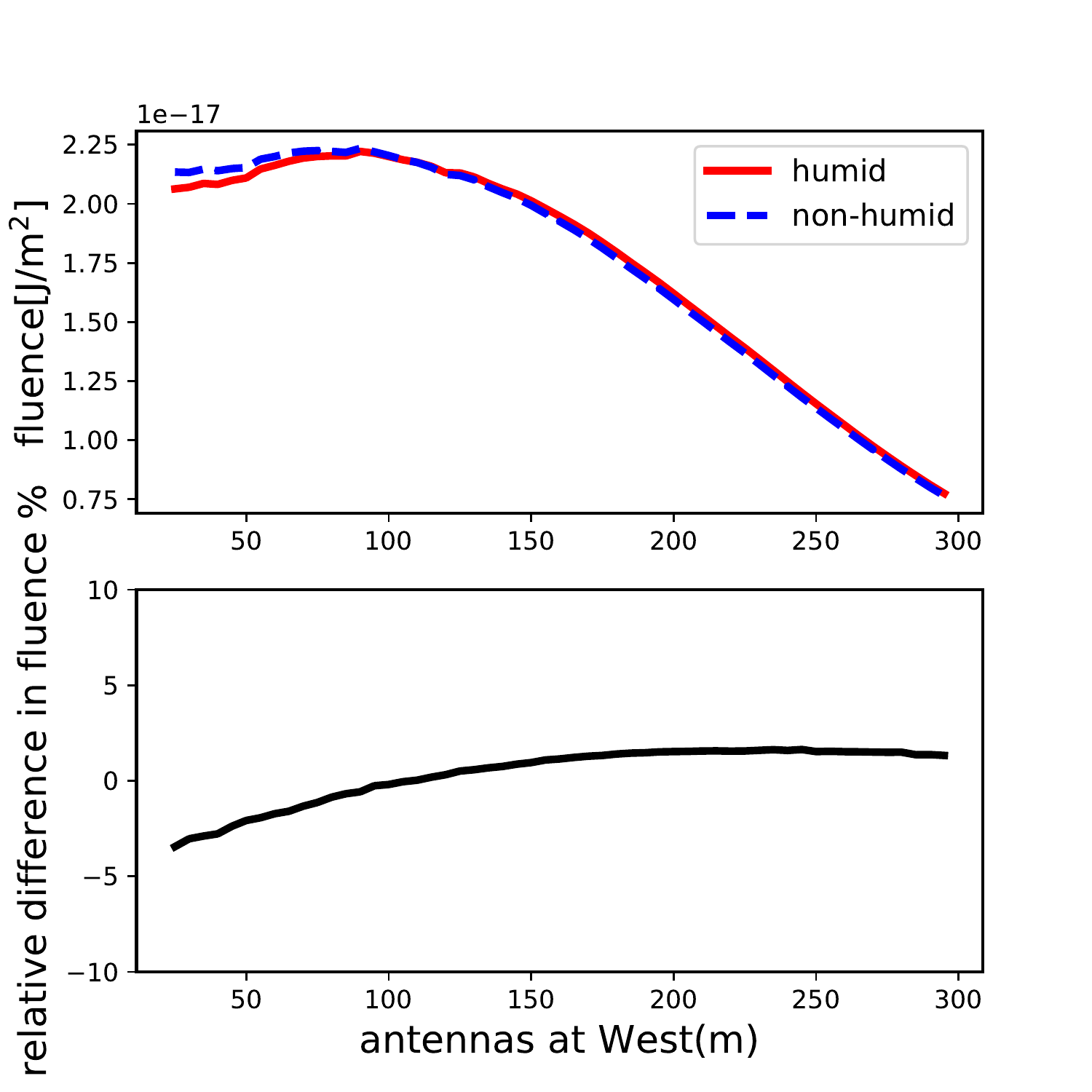}
\includegraphics[width=0.515\linewidth]{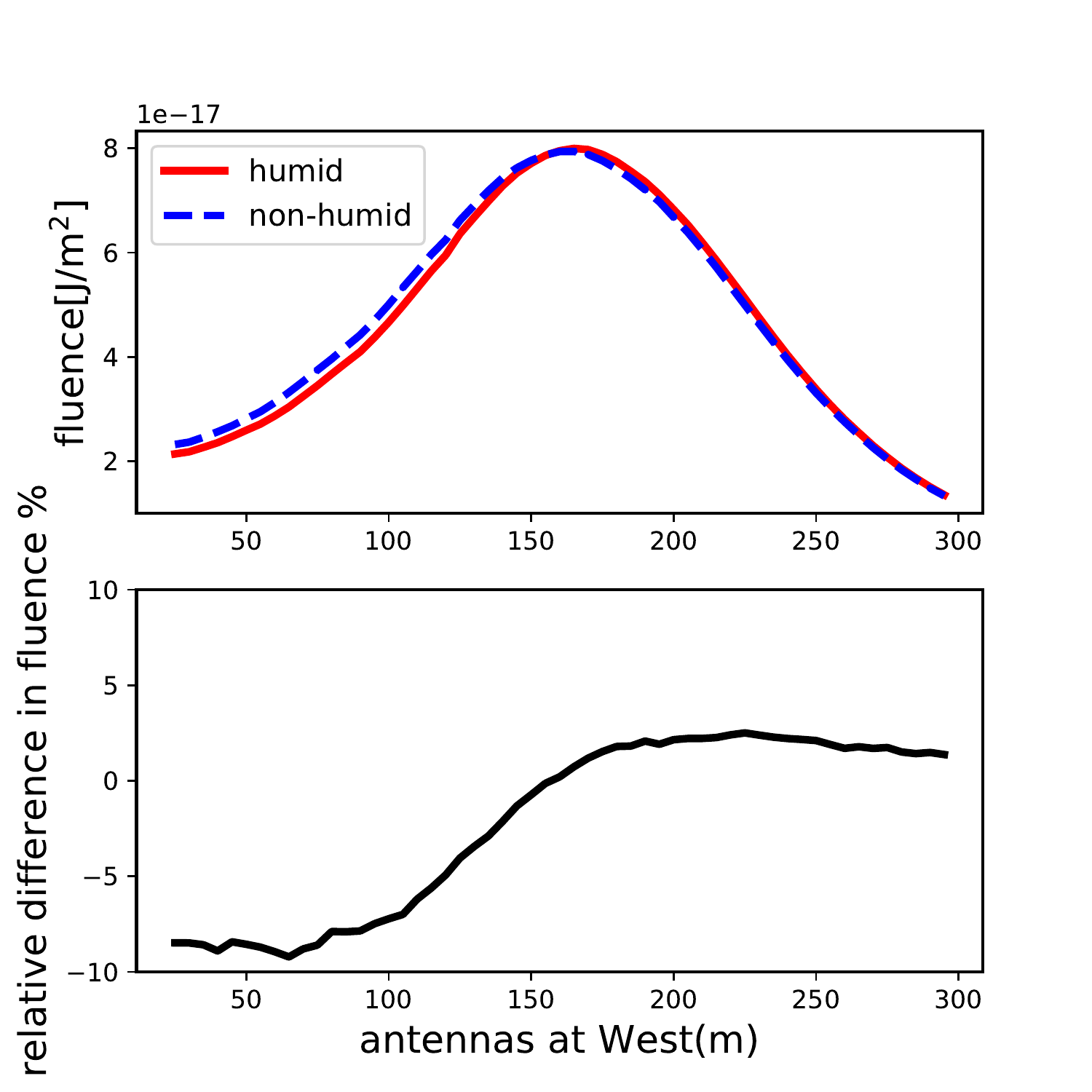}
\caption{Radio lateral distribution profiles for a $10^{17}$ eV proton shower coming from zenith 45$^\circ$  with true $\Xmaxmath =~593\,\mathrm{g/cm^{2}}$. Observers are located
to the west of the shower core. \textbf{Left}: low frequency band between $30$-$80$\,MHz. \textbf{Right}: high frequency band between
$50$-$350$\,MHz. The upper panel shows the LDF of total fluence for the humid and non-humid sets,  the lower panel shows the relative difference between these two. }
\label{cheren}
\end{figure}

\begin{figure}[h!]
\centerline{\includegraphics[width=0.83\textwidth]{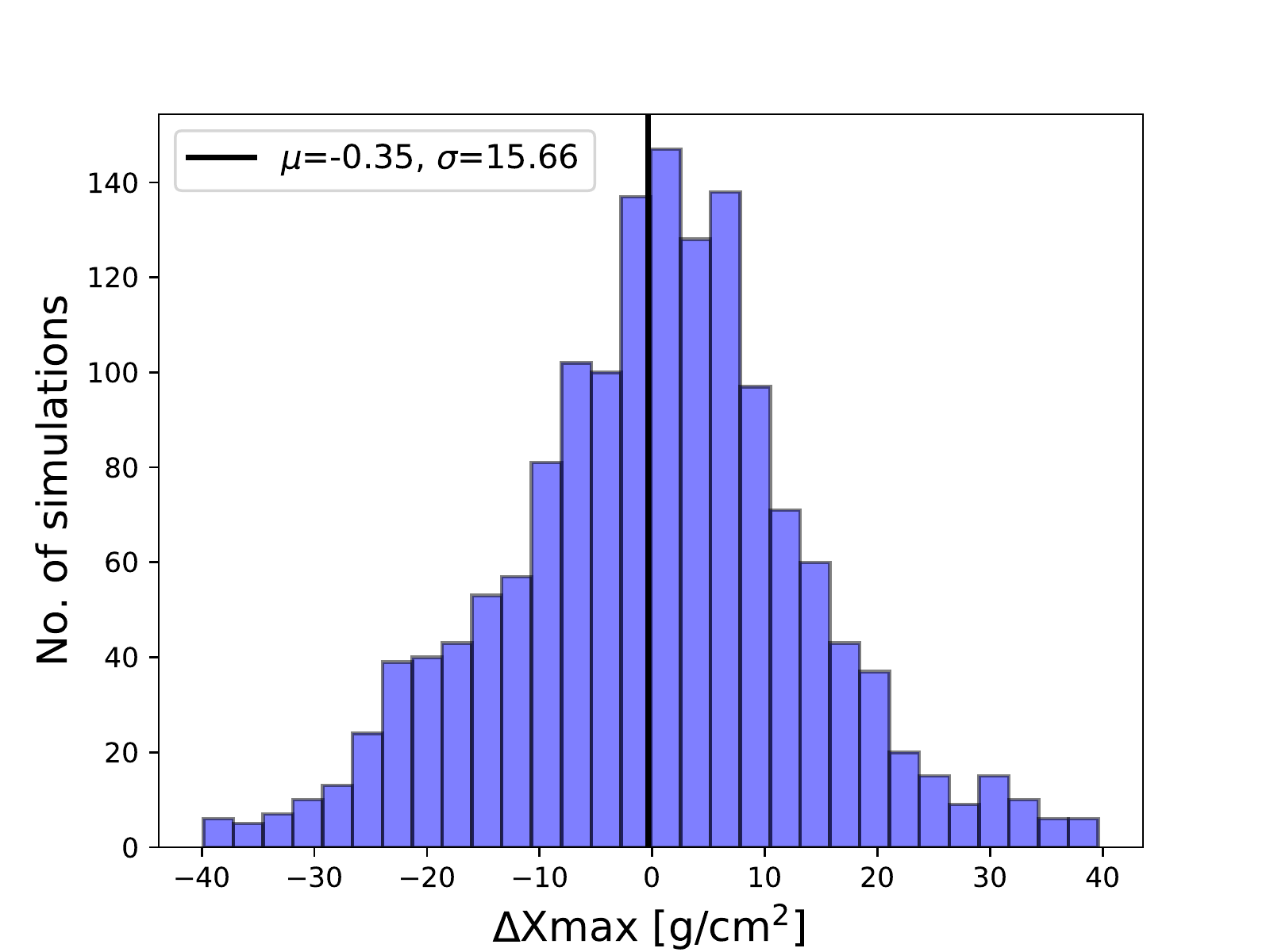}}
\centerline{\includegraphics[width=0.83\textwidth]{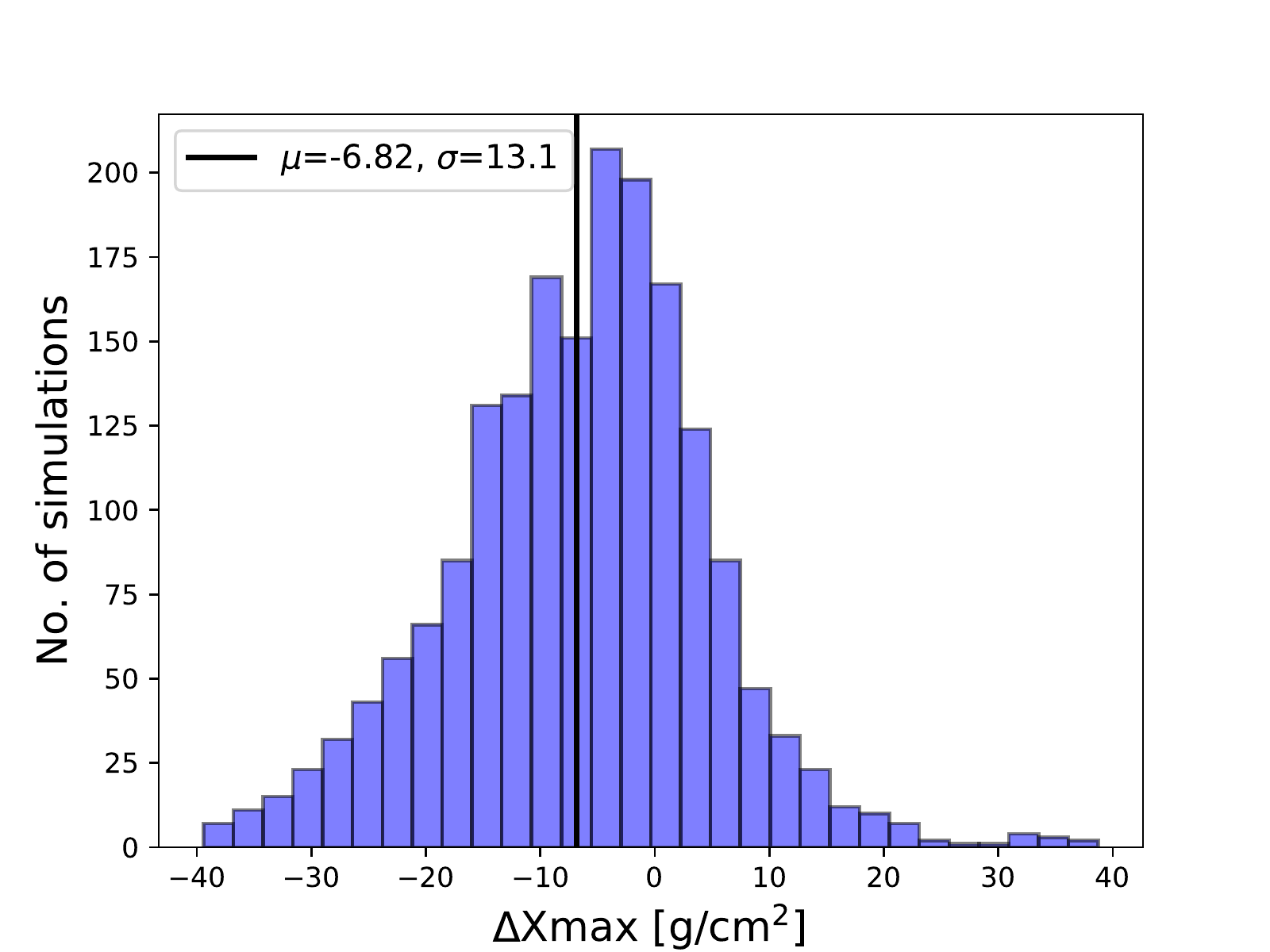}}
\caption{Histogram for the $\mathrm{\Delta{X_{\max}}}=X_\mathrm{{reco}}{-}\,X_\mathrm{{real}}$ between the reconstructed and true 
value of the \Xmax obtained from the Monte Carlo study between the humid and non-humid simulation sets. \textbf{Top}: for the low frequency band of $30$-$80$ MHz. 
\textbf{Bottom}: for the high frequency band of $50$-$350$ MHz. The shift in the \Xmax is significant at 2$\sigma$ level. }
\label{histxmax}
\end{figure}
To investigate the effect of humidity on \Xmax measurements we have performed a Monte Carlo comparison study between two sets of simulations that
deals with the atmospheres in a similar way as described in the beginning of this section. 
For each of theses cases we have used a set of 40 simulated events with different energy,
zenith and azimuth angles. Each of these showers have an ensemble of proton and iron initiated showers selected with CONEX.
One shower from the set with realistic humidity is taken as reference and all the simulated showers from the set 
with zero humidity are used to perform the reconstruction. This yields a reconstructed
$X_\mathrm{{reco}}$ that can be compared to the actual $X_\mathrm{{real}}$ of the reference shower. 
The same method is repeated for all the showers in the humid set.
The difference
$X_\mathrm{{reco}}{-}\,X_\mathrm{{real}}$ estimates the effect of humidity on reconstructed \Xmax. 
Results are shown in Fig.\,\ref{histxmax}.
  We do not observe any significant shift in \Xmax in this study. 
This indicates that these effects are most likely smaller
than the overall resolution in reconstructed \Xmax in the LOFAR frequency band. We also performed the same study in a higher frequency
band between 50 and 350 MHz. There, an overall shift of 6.8 $\mathrm{g/cm^{2}}$ in the reconstructed \Xmax was observed.

\section{Conclusions}

Atmospheric effects are important for the precise
reconstruction of \Xmax with the radio technique. 
We have developed a tool that extracts atmospheric  
parameters using GDAS for a given time and location. Using this 
we have implemented a time-dependent parameterized density  and refractive index profiles
in CORSIKA/CoREAS. This code is flexible and can be used
for other air shower experiments as well. 
Using LOFAR data, we demonstrated the importance of using full GDAS-based atmospheres instead of a linear geometrical correction to the US standard
atmosphere. While this correction is sufficient for the bulk of the events, it becomes indispensable for extreme values for the air pressure.
When the air pressure at ground level differs by less than 10 hPa from the US standard atmosphere value, the reconstructed \Xmax value 
including the correction agrees to the full GDAS-based reconstruction value within $2\,\mathrm{g/cm^{2}}$. However, for periods of very 
low air pressure, this difference grows significantly up to $15\,\mathrm{g/cm^{2}}$.  
Effects of humidity on the energy fluence and reconstructed \Xmax were probed;
effects become more prominent for higher frequency bands. This could be important for the high precision \Xmax 
measurements for the cosmic-ray detection with the SKA experiment \cite{gdas:ska}.

\FloatBarrier

\begin{footnotesize}
\subsection*{Acknowledgements}
We sincerely thank Tanguy Pierog for assistance regarding the implementation of our work in CORSIKA modules.

\begin{tocless}
\bibliographystylegdas{unsrt}
\bibliographygdas{LOFARbib}
\end{tocless}

\end{footnotesize}

\newpage


\savepagenumber
\part{Influence of atmospheric electric fields on radio
emission from air showers}

\title{Influence of atmospheric electric fields on radio
emission from air showers}

\author{\speaker{O. Scholten}\!\!, G. Trinh for the LOFAR CRKSP\\
		E-mail: \email{O.Scholten@rug.nl}}

\addauthor{U. Ebert\\Centrum Wiskunde\,\&\,Informatica, Science Park 123, 1098\,XG Amsterdam, The Netherlands\\ 
	Eindhoven University of Technology (TU/e), PO Box 513, 5600\,MB Eindhoven, The Netherlands
}

\abstract{We show that atmospheric electric fields as they exist in thunderclouds can strongly affect the radio emission from cosmic air showers. We also show, using data from LOFAR, that from the measured radio footprint of cosmic-ray air showers, i.e., intensity, linear and circular polarization at various distances from the shower core, one can determine the direction and strength of the electric field as function of height along the path of the cosmic ray. This method can be regarded as tomography of thundercloud electric fields using cosmic rays as probes. We will present an analysis of selected events measured during thunderstorm conditions in the period from December 2011 till August 2014. The fields we extract are consistent with the generally accepted charge structure in thunderclouds consisting out of three charge layers.
}

\begin{titlingpage}
\maketitle
\end{titlingpage}
\restorepagenumber
\addtocounter{page}{1} 

\section{Introduction}

Extensive air showers (EAS) induce electric currents, a longitudinal current arising from the charge excess in the shower front and a transverse current arising from the action of the Lorentz force due to the magnetic field of Earth. These currents emit radio waves since their strength changes as function of height in the atmosphere because of the changing number of particles in the EAS~\cite{efields:Wer08,efields:Sch13,efields:Hue16}. In addition, since the showers proceed faster than the propagation velocity of light in air, there is a contribution from the Cherenkov effects~\cite{efields:Vries11,efields:Wer12}. The emitted radio pulse has proven to be very useful for an efficient determination of the properties of the cosmic ray that initiated the EAS~\cite{efields:Bui14,efields:Bui16}. This determination of the shower properties relies on a good understanding of the emission mechanisms. In \secref{Mechanism} we discuss an additional mechanism that contributes to radio emission resulting from the currents induced by atmospheric electric fields in charged clouds~\cite{efields:Tri16}. Atmospheric electric fields may induce rather different intensity patterns, but their contribution can most easily be recognized from abnormal polarization footprints as shown in \secref{Effects}. In turn it is shown in \secref{Efields} that the radio footprint (intensity and polarization) can be used as a tool to study electric fields in clouds.

\section{Radio emission due to atmospheric electric fields}\seclab{Mechanism}

It is well known that lightning is driven by strong electric fields in large Cumulonimbus clouds. These electric fields result from a separation of charges due to strong convection flows of ice particles in clouds~\cite{efields:Dei08}. It is less well known that strong electric fields may also be present in more ordinary clouds.

When an EAS is developing in a region with an atmospheric electric field the electrons and positrons in the shower front experience an electric force in addition to the geomagnetic Lorentz force. Even for  moderate electric fields of 10~kV/m the electric force is larger than the geomagnetic one. In considering the effect of the electric field one should distinguish between the component perpendicular to the shower axis and the one parallel to it~\cite{efields:Tri16}. The effect of the perpendicular component is to induce a transverse current along its direction which in general differs from that of the Lorentz force. This will reflect in the direction of the dominant polarization of the radio emission. The parallel component has a very small effect. The reason for this is that the gain/loss of energy due to this component is negligible for the energetic particles in the shower. Only lower energy particles will be affected, but due to a lack of coherence, as these are found at large distances behind the shower front, this will hardly affect the radio emission in the frequency regime above 30~MHz where most of the observations are made. In addition, effects are balanced since a field that accelerates electrons will decelerate positrons. In leading order the total number of charged particles is thus not affected.

An interesting aspect to note is that when the EAS passes through a charge layer in a cloud there is a change in direction of the force acting on the particles in the shower front. As a result the direction of the induced currents changes as function of height. This gives rise to many interesting phenomena, such as constructive and destructive interference of radiation coming from different heights when the currents flip sign. When the currents at different heights are at an angle the received radio pulse will show a strong circular polarization~\cite{efields:Sche15,efields:Tri17,efields:Trinh18}.

\section{Radio footprints}\seclab{Effects}

To analyze the radio footprint we use the real-valued Stokes parameters, defined as~\cite{efields:Sche15}
\begin{eqnarray}
\eqlab{Stokes}
I &=& \frac{1}{n}\sum_{i=0}^{n-1}\left(\left |\varepsilon_{i,\vB}  \right |^2+\left |\varepsilon_{i,\vvB}  \right |^2\right),
\\
Q &=&  \frac{1}{n}\sum_{i=0}^{n-1}\left(\left |\varepsilon_{i,\vB}  \right |^2-\left |\varepsilon_{i,\vvB}  \right |^2\right),
\\
U +iV&=& \frac{2}{n}\sum_{i=0}^{n-1}\left(\varepsilon_{i,\vB}\,\varepsilon_{i,\vvB}^*\right) \;,
\end{eqnarray}
where $\varepsilon_i =  S_i+i\hat{S}_i$ is the complex-valued radio signal, where $\hat{S}_i$ is sample $i$ of the Hilbert transform of $S$. These Stokes parameters can be evaluated for each antenna. Stokes $I$ corresponds to the intensity of the radio signal, while $Q$ is the intensity difference between the polarized intensity in the $\vB$ direction and the $\vvB$ direction. For a fair-weather shower one thus expects $Q/I\approx 1$ since the signal is dominated by geomagnetic emission. Stokes $U$ is the difference in intensity between the polarization directions at $45^\circ$ and $-45^\circ$ with respect to the $\vB$ axis. For a fair weather event a non-zero value is due to the charge-excess contribution and the sign (and magnitude) depends on the azimuthal angle of the antenna with respect to the $\vB$-axis. Stokes $V$ shows the circular polarization. As argued in Ref.~\cite{efields:Scho16}, the circular polarization in fair-weather showers results from a slight emission-time difference between charge-excess and geomagnetic radiation and sign and magnitude depend on the azimuth angle of the antenna, similar as for $V/I$. An example of the Stokes parameters for a fair-weather shower can be found in Ref.~\cite{efields:Scho16}.

For events recorded at LOFAR~\cite{efields:Haa13} while there were clouds overhead with (apparently) strong electric fields we find completely different footprint signatures. Two examples are given as events A \& B in \figref{83685143} and \figref{83685776}. Both of these events were recorded on August 26, 2012 where event A was measured at 13:52:23~UTC and event B at 14:02:56~UTC within a 10 minute time span. At the time of observation there was lightning activity in the vicinity of the LOFAR core. It should be noted that clear non-fair-weather radio footprints have also been measured while there was no lightning activity within 200~km from the core within 12 hours before or after the event.

\begin{figure}[h]
\centering
	\includegraphics[width=1.023\textwidth]{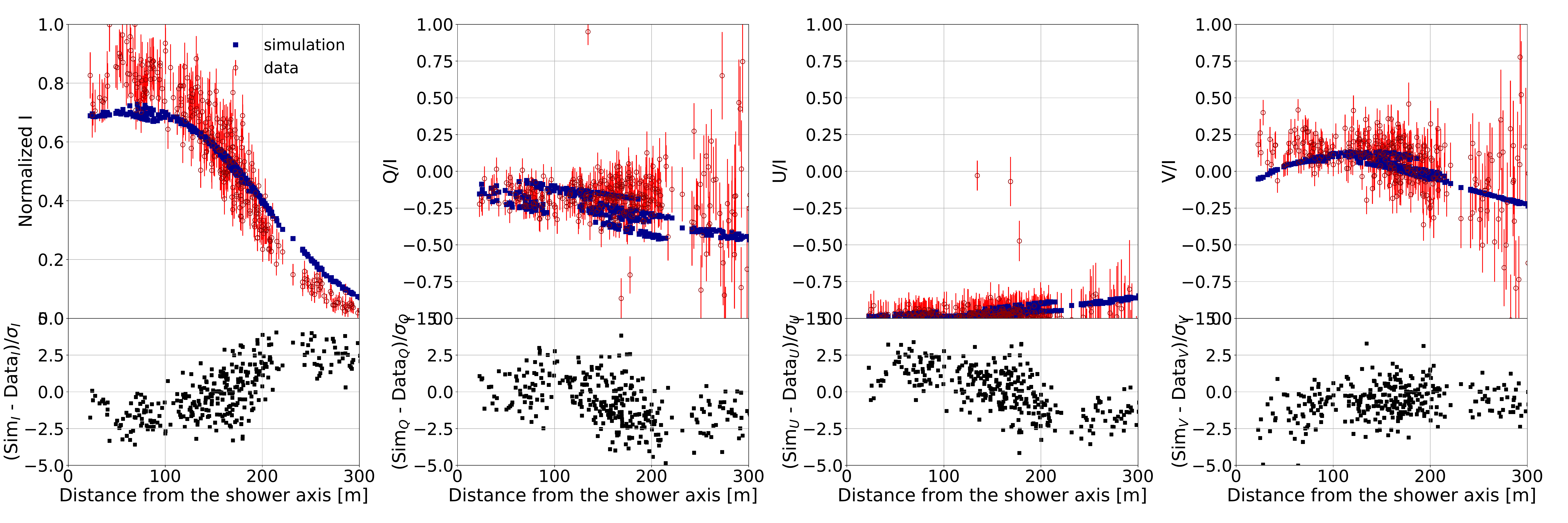}
\caption{The results for normalized Stokes parameters (filled blue dots) calculated with CoREAS, using the field configuration given in \tabref{field-EveAB} are compared to LOFAR data (open red circles) for event A. Bottom panels show the difference between calculation and data normalized by $\sigma$, the one standard deviation error.} \figlab{83685143}
\end{figure}

\begin{figure}[h]
\centering
	\includegraphics[width=1.023\textwidth]{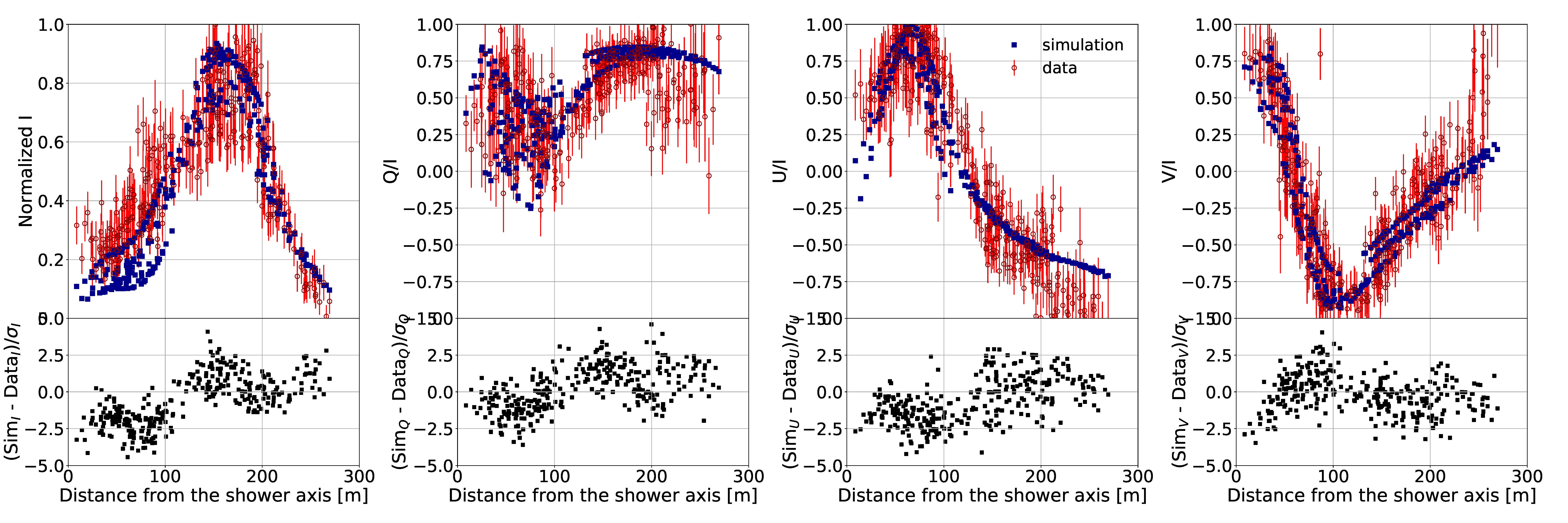}
\caption{Same as \figref{83685143} for event B.}
\figlab{83685776}
\end{figure}

The intensity and circular polarization patterns observed in event A, see red points in \figref{83685143}, resemble those of a fair weather event, however the linear polarization is rather different. Instead of unity we find here $Q/I\approx 0$ which means that the polarization vector is making a $\pm 45^\circ$ angle with the $\vB$ direction. Since $U/I\approx -1$ the angle is actually $-45^\circ$. The fact that the circular polarization is small, $V/I\approx 0$, implies that the fields in the different layers are mostly oriented along the same direction as there should not be a net rotation angle in the fields. The blue points in \figref{83685143} show the results of a CoREAS calculation~\cite{efields:Hue12} using the three-layered structure for the atmospheric electric field as specified in \tabref{field-EveAB}. For the true force acting on the electrons and positrons the geomagnetic force has to be added, resulting is a net force in the $-45^\circ$ plane.

The radio-intensity pattern for event B, occurring only 10~minutes later, is given by the red points in \figref{83685776}. This shows a clear ring-structure in intensity with a diameter of close to 200~m. This is a clear indication of a strong destructive interference between different layers. The strong circular polarization near the core is evidence that the field orientations have a definite twist. From the values of the atmospheric electric field given in \tabref{field-EveAB} one can see that this is indeed the case.

\begin{table}[h]
\begin{center}
    \begin{tabular}{ |r || c| c | c || c | c | c || }
    \hline
   Event  & \multicolumn{3}{c||}{A} & \multicolumn{3}{c||}{B}  \\ \hline
 Energy (eV)  & \multicolumn{3}{c||}{$4.4\times 10^{16}$} & \multicolumn{3}{c||}{$3.9\times 10^{16}$} \\ \hline
    Layer & $h$ & $E$ & $\alpha$ & $h$ & $E$ & $\alpha$ \\ \hline
          1&    9.1 & 57 & -63 &    5.8 & 30 & -29      \\
          2&    4.0 &  3 & 206 &    3.4 & 83 & 180      \\
          3&    1.2 &  4 & -20 &    1.7 & 13 &  30      \\
    \hline
  \Xmax\ (g/cm$^2$) &\multicolumn{3}{c||}{550}  & \multicolumn{3}{c||}{650}\\
  \Xmax\ (km)  & \multicolumn{3}{c||}{ 5.6} & \multicolumn{3}{c||}{4.1} \\
\hline
 $\chi^2_{3D}$ &\multicolumn{3}{c||}{1.08}  & \multicolumn{3}{c||}{2.11} \\
\hline
  $\chi^2_C$ &\multicolumn{3}{c|}{2.02}  &\multicolumn{3}{c|}{1.95}\\
\hline
  $f_r$ &\multicolumn{3}{c|}{3}  &\multicolumn{3}{c|}{8}\\
\hline
    \end{tabular}
\end{center}
 \caption{The values of the parameters describing the structure of the atmospheric electric field assuming three-layers. For each layer the top-height $h$ of the layer is given [km], the strength $E$ of the electric field [kV/m], and the angle of the field with the $\vB$ direction. Also given are the values for the chi-square for the MGMR3D calculation ($\chi^2_{3D}$) and that for CoREAS ($\chi^2_C$) using the same field and almost the same \Xmax. The normalization factor for the intensity of the radio signal is given by $f_r$.}
   \tablab{field-EveAB}
\end{table}

\begin{table}[h]
\begin{center}
    \begin{tabular}{ |r || c| c | c || c | c | c || }
    \hline
   Event  & \multicolumn{3}{c||}{A} & \multicolumn{3}{c||}{B}  \\ \hline
   Time  26/08/2012 & \multicolumn{3}{c||}{ 13:52:23 UTC} & \multicolumn{3}{c||}{14:02:56 UTC}  \\ \hline
   ($\theta , \phi$)& \multicolumn{3}{c||}{ (23$^\circ$, 144$^\circ$)} & \multicolumn{3}{c||}{(18$^\circ$, 310$^\circ$)}  \\ \hline
 Energy [eV]  & \multicolumn{3}{c||}{$4.4\times 10^{16}$} & \multicolumn{3}{c||}{$3.9\times 10^{16}$} \\ \hline
          & $h$ & $E_\vz$ & $E_\vvz$ & $h$ & $E_\vz$ & $E_\vvz$ \\
    Layer & [km] & [kV/m] & [kV/m]  &    [km] & [kV/m] & [kV/m] \\ \hline
          1&    9.1 & -47.3 & 32.3  &    5.8 & 14.1 &  26.2   \\
          2&    4.0 &   1.9 &  2.7  &    3.4 &  1.0 & -82.5   \\
          3&    1.2 &  -4.3 & -0.6  &    1.7 & -6.9 &  11.6   \\
\hline
    \end{tabular}
\end{center}
 \caption{The values of the parameters describing the structure of the atmospheric electric field assuming three-layers. For each layer the top-height $h$ of the layer is given [km], the strength $E$ of the electric field [kV/m], and the angle of the field with the $\vB$ direction. Also given are the values for the chi-square for the MGMR3D calculation ($\chi^2_{3D}$) and that for CoREAS ($\chi^2_C$) using the same field and almost the same \Xmax. The normalization factor for the intensity of the radio signal is given by $f_r$.}
   \tablab{field-EveAB2}
\end{table}

\section{Extracting atmospheric electric fields}\seclab{Efields}

From the values given in \tabref{field-EveAB} one can see that the structure of the atmospheric electric field can be rather complicated. On the basis of the general charging mechanism in clouds one expects around the freezing level a positively charged layer, at ${-}10^\circ$~C the main negative layer and near the top of the cloud another positive layer~\cite{efields:Dei08}. Based on this we assume a three-layered structure. The problem we are now facing is that for a measured radio footprint we have to search for the nine parameters that define such a three-layered configuration. This is too many parameters to deal with using CoREAS~\cite{efields:Hue12} because of running time and the stochastic nature of the calculations. To solve this problem we have developed a semi-analytic code MGMR3D~\cite{efields:Sch18,efields:Sch18AR} that can calculate a complete radio footprint in just a few seconds. For this reason it can be used in a chi-square optimization. MGMR3D uses a simplified parametrization for the structure of the air shower and thus yields only an approximation to the full-scale calculation. To be sure about the configuration we thus perform a CoREAS~\cite{efields:Hue12} calculation where the atmospheric electric fields (found with MGMR3D) are implemented through the EFIELD option~\cite{efields:Bui10}.

The first step in the procedure for constructing the atmospheric electric fields it thus to fit the measured cosmic-ray footprint using MGMR3D. This may require several attempts to ensure that one is not stuck in a local minimum. It also turns out that the value of \Xmax\ is strongly correlated with the field configuration making it impossible to fit this in conjunction with the field configuration. We thus perform the fit for several fixed values for \Xmax. In a second stage the obtained field configuration is used is a CoREAS calculation where also the measured scintillator signal is taken into account. The final selection is based on the agreement between CoREAS and the data as well as the absolute intensity of the radio emission (expressed by $f_r$ where $f_r=1$ implies good agreement and the energy of the shower is determined from the scintillator signal).

The thus obtained values are given in \tabref{field-EveAB} and the quality of the CoREAS results can be judged from the value of $\chi^2_C$ in the table or even better from \figref{83685143} and \figref{83685776}. One observes that in spite of rather complicated interferences between the emission from different heights the field configuration obtained from the MGMR3D calculations gives satisfactory results when used in CoREAS.

\subsection{E-Field Tomography}

The radio footprint is primarily sensitive to the field perpendicular to the shower axis. Thus, when two showers coming at different angles are measured within a short time span, one should be able to reconstruct the complete field. This we call E-Field Tomography. For this to work the basic assumption is that the fields does not change during the time between the showers as well as the the fields are the same along the tracks of the showers which do not coincide. It is difficult to give general scales for time and distances since this will depend on cloud size (typically 10~km, but varies much), wind speeds. The tomography method also offers a test for the change in the field.

To perform tomography we consider two showers $i$ and $j$ with shower axes given by $\mathbf{v}_i$ and $\mathbf{v}_j$. For each of these the perpendicular components of the electric fields, $\mathbf{E}_{\perp i}$ and $\mathbf{E}_{\perp j}$, can be determined. Assuming that the complete field $\mathbf{E}$ is the same for the two showers, we may write
\begin{equation}
\mathbf{E}_{\perp i} + E_{i\parallel} \mathbf{v}_i  =
\mathbf{E} = \mathbf{E}_{\perp j} + E_{j\parallel} \mathbf{v}_j \,,
\eqlab{vecE}
\end{equation}
where $E_{i\parallel}$ and $E_{j\parallel}$ are the parallel components of the fields.
Taking the dot product of \eqref{vecE} with $e_{\mathbf{v}_i \times \mathbf{v}_j}$, $\mathbf{v}_i$ and $\mathbf{v}_j$ the following three equations are obtained,
\begin{equation}
\mathbf{E}_{\perp i}\cdot (e_{\mathbf{v}_i \times \mathbf{v}_j}) = \mathbf{E}_{\perp j} \cdot (e_{\mathbf{v}_i \times \mathbf{v}_j})\,,
\eqlab{check_consist}
\end{equation}
and
\begin{eqnarray}
E_{\parallel i}& =& \frac{\mathbf{E}_{\perp j} \cdot \mathbf{v}_i + (\mathbf{v}_i\cdot\mathbf{v}_j) (\mathbf{E}_{\perp i}\cdot\mathbf{v}_j)} {1-(\mathbf{v}_i\cdot\mathbf{v}_j)^2} \nonumber \\
E_{\parallel j}& =& \frac{\mathbf{E}_{\perp i} \cdot \mathbf{v}_j + (\mathbf{v}_i\cdot\mathbf{v}_j) (\mathbf{E}_{\perp j}\cdot\mathbf{v}_i)} {1-(\mathbf{v}_i\cdot\mathbf{v}_j)^2} \,.
\eqlab{Epara}
\end{eqnarray}
\eqref{check_consist} can be considered as a consistency check i.e.\ that both measurements yield the same component of the field in the $e_{\mathbf{v}_i \times \mathbf{v}_j}$-direction which is perpendicular to both showers.  \eqref{Epara} allows the construction of the as yet missing parallel component of the field. This procedure can be applied to each layer separately.

\begin {table}[h!]
\begin{center}
 \begin{tabular}{|r @{ -- } l||c|c||c|c||c||}
 \hline
  event A & event B & $\mathbf{E}_A\cdot(e_{\mathbf{v}_A \times \mathbf{v}_B})$ & $\mathbf{E}_B\cdot(e_{\mathbf{v}_A \times \mathbf{v}_B})$ & \ \ \ $\mathbf{E}_z$\ \ \  & \ \ \ $\mathbf{E}_z$ \ \ \ & height\\
\hline
\hline Top     & Top     &     43 & 10  & -94 & -95 & $\approx$8 km\\
\hline Middle  & Middle  &     -2 & 13  & 113 & 114 & 3.5 km\\
\hline Bottom  & Bottom  &      4 & -9  & -15 & -15 & 1.5 km\\
\hline
\end{tabular}
\caption{Checking the consistency of electric fields extracted from events A and B (see \eqref{check_consist}). The quoted values are in [kV/m]. The parallel components of the fields have been determined using \eqref{Epara} when reconstructing the vertical, $z$, component of the field.}
\tablab{E_compr}
\end{center}
\end {table}

In \tabref{E_compr} we show the results when tomography is applied to events A and B. One sees that the consistency condition is obeyed to a reasonable extent. The table also gives the extracted vertical component of the field. This shows that the layer at an height of 4.0 or 3.4~km is strongly negatively charged. This is also close to the height of the ${-}10^\circ$-isotherm  as expected. The $0^\circ$-isotherm lies at 2.5~km at the time of this event which is a bit higher than the height of the lower positive charge layer which we determine from these data. A more extensive discussion of these events will appear in Ref.~\cite{efields:Tri20}

\section{Summary}

We argue that atmospheric electric fields may have a large effect on the radio footprint of cosmic-ray air showers. We have shown the results for a measurement during an active thunderstorm, however strong electric fields have also been observed for heavy rain clouds.

It is shown that from the measured footprint it is possible to reconstruct the component of the electric field that is perpendicular to the shower. A semi-analytic code has been developed to be able to do so efficiently. Tomography can be applied when multiple events are measured within a short time-span. This allows for the reconstruction of the complete field and thus the charge structure in the cloud.

It should be noted that at GRAPES-3 muon telescope located in Ooty, India, strong variations in muons have been measured that have been used to determine potential differences in thunderclouds of 1.3~GV~\cite{efields:Har19}. Also at the Pierre Auger Observatory ring-like structures have been observed in the surface detectors with diameters of the order of a few km that can be explained as due to strong atmospheric electric fields of the order of 500~kV/m extending over a distance of a kilometer~\cite{efields:Col19}.

\begin{tocless}
\bibliographystyleefields{unsrt}
\bibliographyefields{LOFARbib}
\end{tocless}

\newpage


\savepagenumber
\part{Extension of the LOFAR Radboud Air Shower Array}

\title{Extension of the LOFAR Radboud Air Shower Array}

\author{\speaker{K. Mulrey}\!\! for the LOFAR CRKSP\\
		E-mail: \email{kmulrey@vub.ac.be}}

\addauthor{D. Veboric\\ Institut f\"{u}r Kernphysik (IKP), Karlsruhe Institute of Technology (KIT),\\ 
P.O.\,Box 3640, 76021\,Karlsruhe, Germany\\}

\abstract{The LOFAR Radboud Air Shower Array (LORA) is an array of scintillators situated at the core of the LOFAR radio telescope. LORA detects particles from extensive air showers and acts as a trigger for the readout of the LOFAR antennas, which are densely spaced and routinely measure radio emission from air showers around $10^{17}$\,eV.  LORA originally consisted of 20 scintillators. An extension is underway that doubles the number of scintillators and increases the effective area of the array.  This will result in a 45\% increase in the number of triggers from higher energy cosmic rays, which are more likely to produce a strong radio signal. In addition, it will reduce the composition bias inherent in detecting low energy showers. In this contribution we discuss the status of the LORA extension and prospects for the science that can be done with the expanded triggering capabilities and improved calibration of the detector.}

\begin{titlingpage}
\maketitle
\end{titlingpage}
\restorepagenumber
\addtocounter{page}{1}

\section{Introduction}

Cosmic rays interacting in the Earth's atmosphere create particle air showers, where radio emission is generated through the charge-excess effect and the transverse current induced by the magnetic field of the Earth.  The pattern of radiation on the ground contains information about the primary cosmic ray~\cite{lora:huege2016}, which is used to reconstruct features including energy, geometry, and composition with high precision \cite{lora:buitink2014}.  The LOw Frequency ARray (LOFAR) is especially well suited to study radio emission from cosmic rays because of its dense antenna spacing \cite{lora:schellart2013,lora:LOFAR}.  

The LOfar Radboud Air shower array (LORA) consists of plastic scintillators situated at the LOFAR core.  It acts as a trigger for the LOFAR antennas and provides information about the primary cosmic ray based on particle data \cite{lora:thoudam2014}.  The LORA scintillators have been recycled from the KASCADE experiment~\cite{lora:KASCADE_NIM}.  LORA provides a trigger for LOFAR antenna readout when a given number of scintillators have detected an air shower, with the trigger level for each scintillator roughly corresponding to the amount of energy deposited by one muon.  In order not to interfere with regular astronomical LOFAR operations, the trigger rate must be limited to about one readout an hour.  

The steeply falling cosmic ray spectrum means that most triggered events are low in energy, around $10^{16}$ eV, where the radio signal is unlikely to be detectable above background noise.  There is also a composition bias introduced by the triggering scheme that has to be taken into account, due to the fact that air showers from lighter primaries are more likely to produce a trigger.  LORA originally consisted of 20 scintillators located on the superterp, the most densely populated region of antennas in the LOFAR core.  In order to increase the number of quality events detected, an expansion of LORA is underway which will add 20 new scintillators to the array.  It will reduce the trigger bias and increase the effective area.  We have also revisited the calibration of the new and existing LORA scintillators in order to reduce the systematic uncertainties in the measurements and better understand biases in triggering on particles.

\section{Expansion}

Currently, only 19\% of LORA triggered events have a detectable cosmic-ray radio signal.  This is because LORA triggers primarily on showers that are below $10^{16.5}$ eV.  Triggering on only high energy showers could be achieved by increasing the number of detectors necessary to form a trigger.  However, this reduces the trigger rate to an unnecessarily low level.  Furthermore, many events that do contain a radio signal suffer from triggering biases introduced by triggering on particles.  Since a primary science result for LOFAR cosmic rays is composition studies, decreasing the composition bias will increase the number of usable events.

In order to design the extension of LORA, a simulation study was carried out to determine the optimal layout for the 20 new scintillators at the LOFAR core using CORSIKA v-7.4387~\cite{lora:corsika}, with QGSJET II-04~\cite{lora:qgsjet}, FLUKA~\cite{lora:fluka}, and GEANT4~\cite{lora:geant4}.  Details of the simulations can be found in~\cite{lora:Mulrey:2017ldx}. Consideration had to be taken to install the new scintillators at existing LOFAR antenna stations where the infrastructure was already in place.  Scintillators were grouped in sets of four.  The resulting layout is shown in the left panel of Fig.\,\ref{fig:map}.  The blue squares indicate the positions of the existing scintillators, and the red squares indicate the positions of the new ones.  The right panel of Fig.\,\ref{fig:map} shows the positioning of the scintillators relative to the antenna fields (note: scintillators are roughly 1m $\times$ 1m and are not shown to scale). 

\begin{figure}[h]
\centering
\includegraphics[scale=0.6]{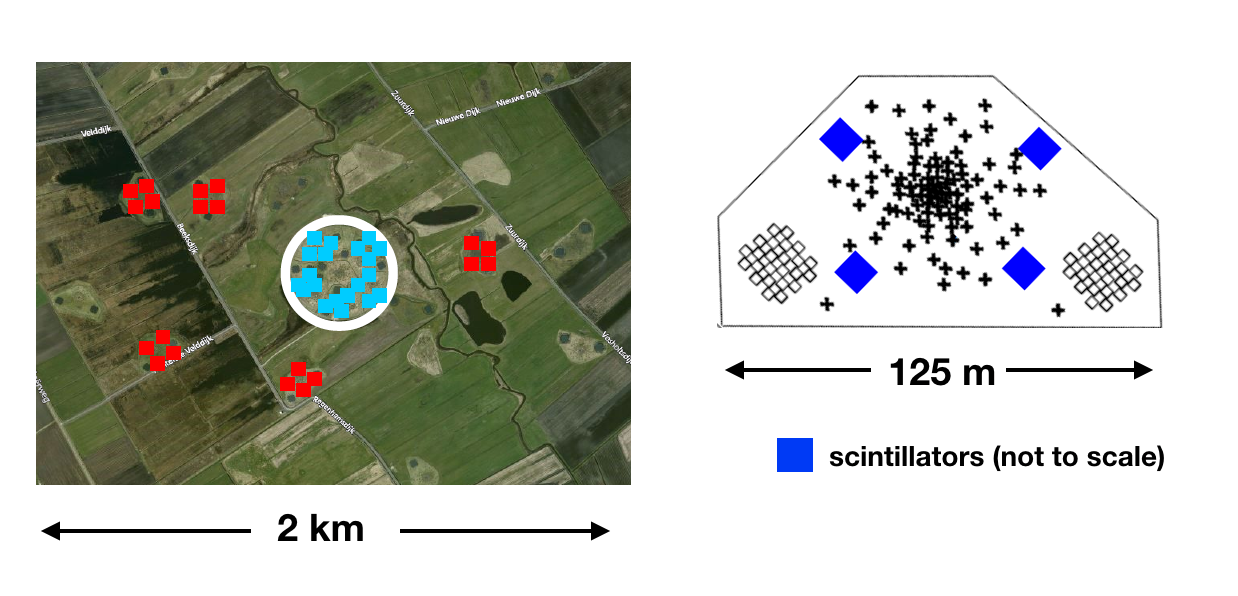}
\caption{Layout of the new LORA stations.  \textbf{Left:}~New scintillator positions are shown in red, and existing ones in blue.  The white circle indicates the position of the superterp, which has the densest antenna spacing.  \textbf{Right:}~Positioning of the scintillators relative to the antenna fields.  Scintillators are not shown to scale.  Crosses indicate low band ($30$-$80$\,MHz) antennas, and boxes indicate high band ($120$-$240$\,MHz).}
\label{fig:map}
\end{figure}

With the added LORA stations, the trigger requirement will be adjusted so that the rate remains at 1~trigger per hour.  This means the minimum number of panels required to form a trigger increases, thereby limiting the number of low energy showers triggering.  The increased effective area also allows for triggering on higher energy showers.  The results of the increased effective area as determined by simulation are shown in the left panel of Fig.\,\ref{fig:sim_results}, where a 45\% increase in usable events is expected.  

A second benefit of adding more scintillators to the array is that the effect of triggering bias is reduced.  Because heavier composition cosmic rays interact further up in the atmosphere, the shower develops higher up, and more particles are attenuated by the time the shower front reaches ground level.  This means that showers initiated by lighter particles have a higher chance of triggering.  This effect is enhanced at higher zenith angles.  In order to eliminate trigger biases, we check on an event-to-event basis that the geometry of the event is such that both proton and iron initiated showers would generate a trigger.  This results in most events below $10^{17}$~eV being discarded. By adding more scintillators, we increase the probability that more showers trigger.   This is shown in the right panel of Fig.\,\ref{fig:sim_results}.  The probability of detecting proton and iron showers with a core within 250~m of the center of the superterp is shown as a function of the number of scintillators required for a trigger.  With the new extension, both primaries trigger at close to 100\% down to $10^{16.5}$~eV for a trigger condition of 13/20 scintillators with signal.  This increases the number of quality events, in addition to the 45\% increase due to the expanded effective area. 

\begin{figure}[h]
\centering
\includegraphics[scale=0.45]{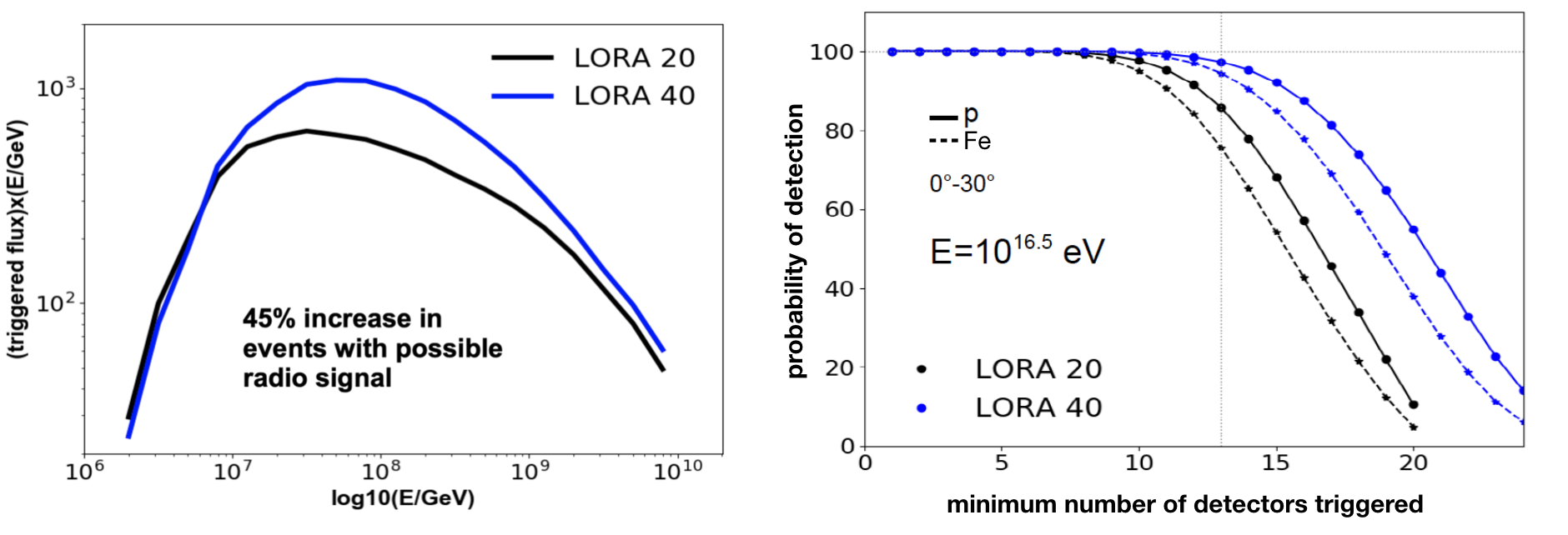}
\caption{\textbf{Left:}~CR flux multiplied by the probability of triggering.  The trigger condition is chosen to keep the rate under 1 event per hour, and so different detector arrangements have different conditions.  \textbf{Right:}~Probability of triggering a minimum number of detectors for a shower of energy $10^{16.5}$~eV within 250~m of the center of the superterp. Solid lines represent proton showers and dashed represent iron showers.}
\label{fig:sim_results}
\end{figure}

The left panel of Fig.\,\ref{fig:cores} shows the locations of shower cores triggered with the original LORA configuration in red, and with the expansion in blue, with trigger conditions such that the trigger rate remains at 1 event per hour (13/20 scintillators for the original array and 14/20 for the extended array).  As well as increasing the effective area for triggering, triggering with cores well outside the superterp will allow the densest antenna area to probe the shower footprint outside the Cherenkov cone, where the emission from all along the shower track is less compressed in time, providing more information about shower development.  An example radio footprint for a $10^{17}$~eV, $45^{\circ}$ air shower in relation to LOFAR low-band antennas is shown in the right panel of Fig.\,\ref{fig:cores}.

\begin{figure}[h]
\centering
\includegraphics[scale=0.45]{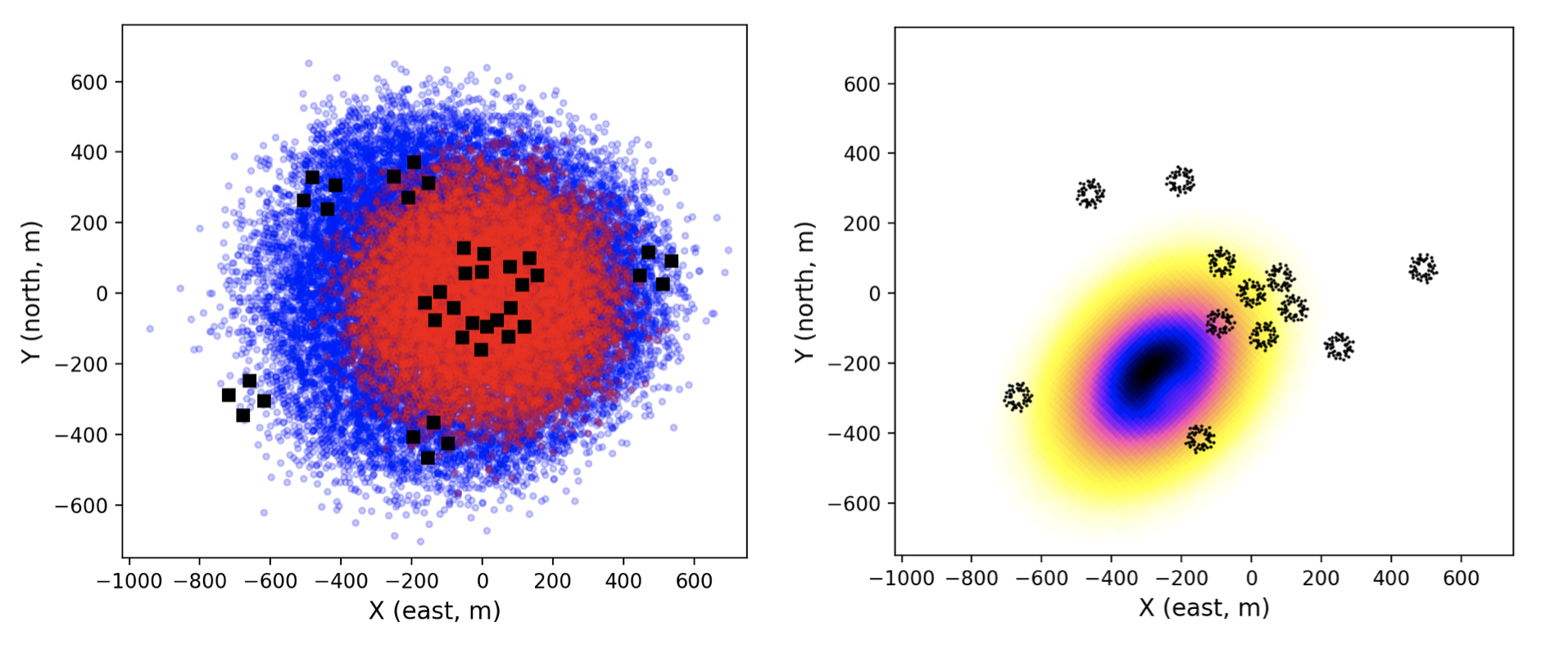}
\caption{\textbf{Left:} Positions of cores of showers triggered with the original 20 scintillators (red) and the new arrangement (blue) for showers with energy $10^{17}$~eV.  The black squares indicate the positions of scintillators.  \textbf{Right:} Sample simulated radio footprint for a shower of $10^{17}$~eV and $45^\circ$~zenith, with positions of the LOFAR antennas. }
\label{fig:cores}
\end{figure}

The first stage of the extension took place in the spring of 2018, when the infrastructure for the new scintillators was installed.
This included laying cabling and adapting the central LOFAR electronics cabinets.  The scintillators were calibrated in the laboratory and installed in the field in spring of 2019. 
Data taking is expected to commence in 
fall of 2019.

\newpage

\section{Improved calibration}

During the LORA expansion we have revisited the calibration of the scintillators in order to reduce systematic uncertainties and better understand biases in the trigger system.  We studied the calibration based on the energy deposited by single charged particles, the spatial dependence of the scintillator response, and reflections in the cabling.

\subsection{Deposited energy from single muons}
It is important to understand the integrated ADC counts of a trace produced by a single muon, which is a proxy for energy deposited in the scintillator.  Here we compare three sets of data: simulated muon energy deposits, muons measured with a LORA scintillator in a muon tracking detector, and muons measured with a LORA scintillator using the electronics that will be used in the LORA extension.

The conversion from integrated ADC counts to deposited energy is done by simulating the detector response to all-sky muons and comparing the expected energy deposit to the measured deposit.  GEANT4 simulations are used, and more details of the simulation procedure can be found in~\cite{lora:thoudam2014}.  The distribution of simulated energy deposits is shown as the orange curve in the right panel of Fig.\,\ref{fig:muon_hist}.  The most probable energy deposit in the scintillator is 6.3~MeV, but the distribution here is scaled so that the most probable value (MPV) is 100 integrated counts to be comparable with the other measurements.  The shape follows the expected Landau distribution.

\begin{figure}[htb]
\centering
\includegraphics[scale=0.42]{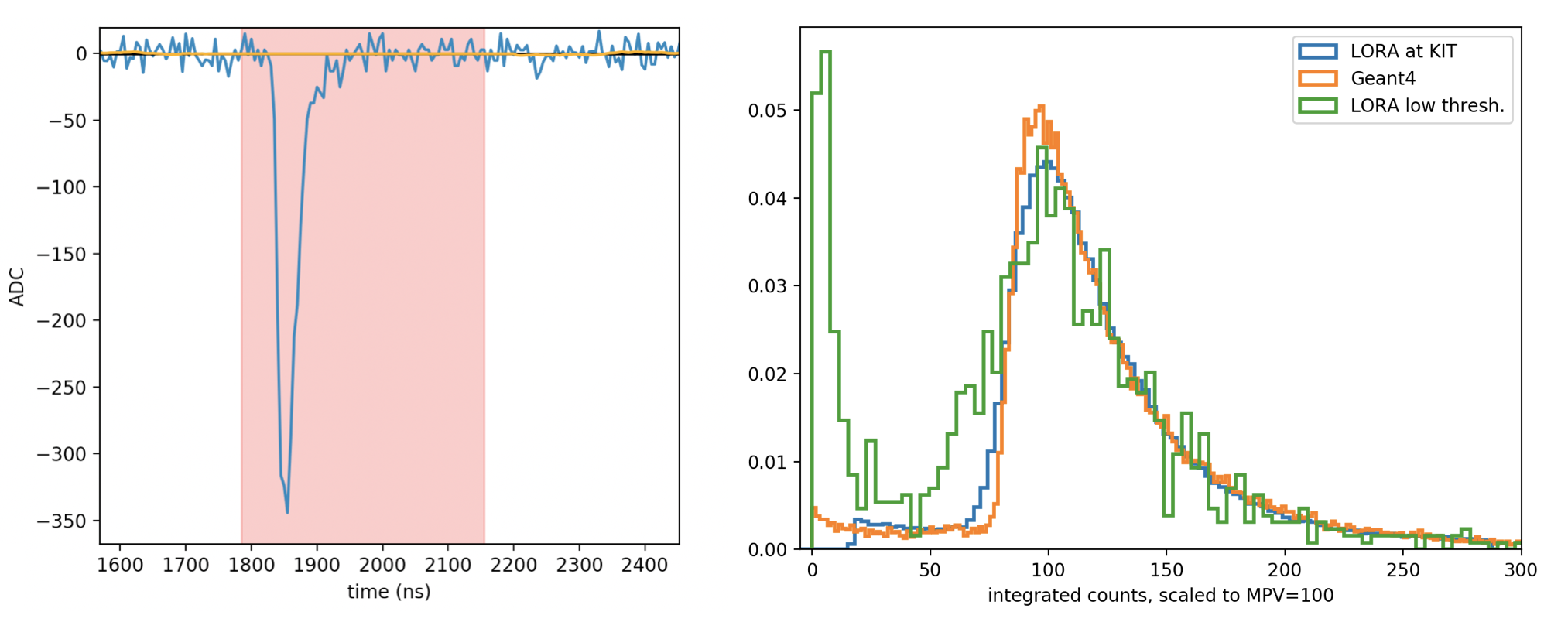}
\caption{\textbf{Left:} Example trace of a single muon in a LORA scintillator.  The red overlay indicates the integration window for determining the total ADC counts, which corresponds to energy deposited in the scintillator.  \textbf{Right:}  Distributions of energy deposits by singly charged particles in the scintillator, scaled so that the most probable value is 100.  The blue line indicates measurements made in the KIT muon tower, orange indicates GEANT4 simulated muons, and green indicates muons measured close to the noise floor.}
\label{fig:muon_hist}
\end{figure}

Four scintillators were sent to Karlsruhe Institute of Technology (KIT) to be measured in a muon tracking detector which was originally used for muon tracking in the KASCADE experiment~\cite{lora:KASCADE_NIM}.  This measurement allowed us to study the distribution of deposited muon energy as a function of position in the scintillator and with high statistics.  The all-sky muon energy deposits are shown as the blue curve in the right panel of Fig.\,\ref{fig:muon_hist}, scaled to MPV=100 counts.  The shape of the distribution matches well with expectation from simulations.

\begin{figure}[t]
\centering
\includegraphics[scale=0.42]{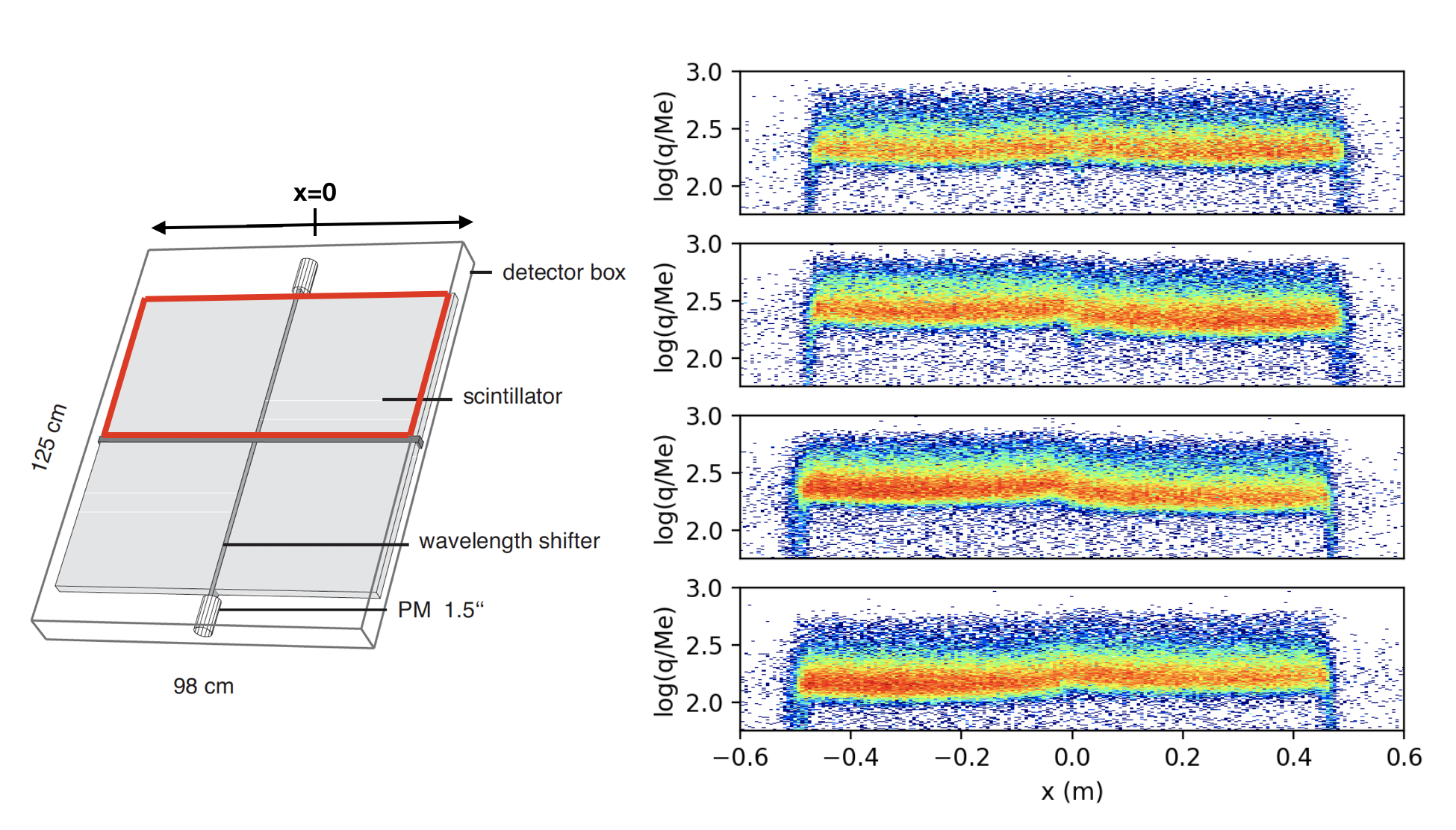}
\caption{\textbf{Left:} Schematic view of the LORA scintillators.  One PMT sees light from two panels joined by a wavelength shifter, indicated by the red box.  Original scintillator schematic adapted from~\protect\cite{lora:KASCADE_NIM}.  \textbf{Right:}~Deposited charge as a function of position along the x-axis of the scintillator, as defined in the left panel.  The four plots are examples of different panel pairs.  The color scale indicates the counts in each bin.}
\label{fig:slices}
\end{figure}

Additionally, we calibrated all the new scintillators by measuring the energy deposits (i.e. integrated ADC\,counts) of single muons with the electronics that will be used in the field.  In order to make use of the full dynamic range of the digitizing electronics, we set the gain so that the signal from a single muon is close to the noise level.  This way we avoid saturation when measuring strong signals from air showers.  A sample muon trace is shown in the left panel of Fig.\,\ref{fig:muon_hist}.  The red overlay shows the integration window, which extends from\,$-70$ to\,$+300\,$ns from the position of the pulse peak.  The resulting distribution is shown in green in the right panel of Fig.\,\ref{fig:muon_hist}, again scaled to MPV=100.  The distribution of deposits from single muons measured close to the noise level is not as clean as the distribution from simulations from the KIT measurements, as the background noise can add constructively or destructively.  However, the shapes of all three distributions agree for deposits above the MPV.  The noise floor is also clearly separable from the muon peak.  With this knowledge, we can determine the correct conversion from measured ADC counts to deposited energy.

\subsection{Spatial dependence of the scintillator response}
In order to find the particle-based energy for a cosmic-ray event, particles from CORSIKA air showers reaching ground level are run through a GEANT4 simulation using a realistic model of the scintillators~\cite{lora:geant4}.  For this reason, it is important to know the spatial dependence of the scintillator responses.  This was measured in the KIT muon tower for 4 scintillators.  The layout of a scintillator is shown in the left panel of Fig.\,\ref{fig:slices}.  Each unit contains two PMTs which see light from two scintillating panels each, connected with a wavelength shifting bar.  The right panel shows the deposited charge as a function of position along the x-axis of the scintillator, as defined in the left panel, for four PMTs.  Besides a deposit offset between some pairs, the distribution of charge along the panel is relatively constant, differing most noticeably with an increase near the wavelength shifter.  This spatial dependence can now be included in simulations.

\subsection{Cable reflections}
The signal carrying coaxial cables from two PMTs are joined using a BNC splitter.  This causes a drop in signal from each PMT, as well as a reflection of the main signal.  An example is shown in Fig.\,\ref{fig:reflections}.  In the left panel, a signal is shown directly from the PMT (top), and after a BNC splitter is used to join the signal from two PMTs (bottom).  Calibration data is taken for each scintillator both with and without reflections.

\begin{figure}[b]
\centering
\includegraphics[scale=0.553]{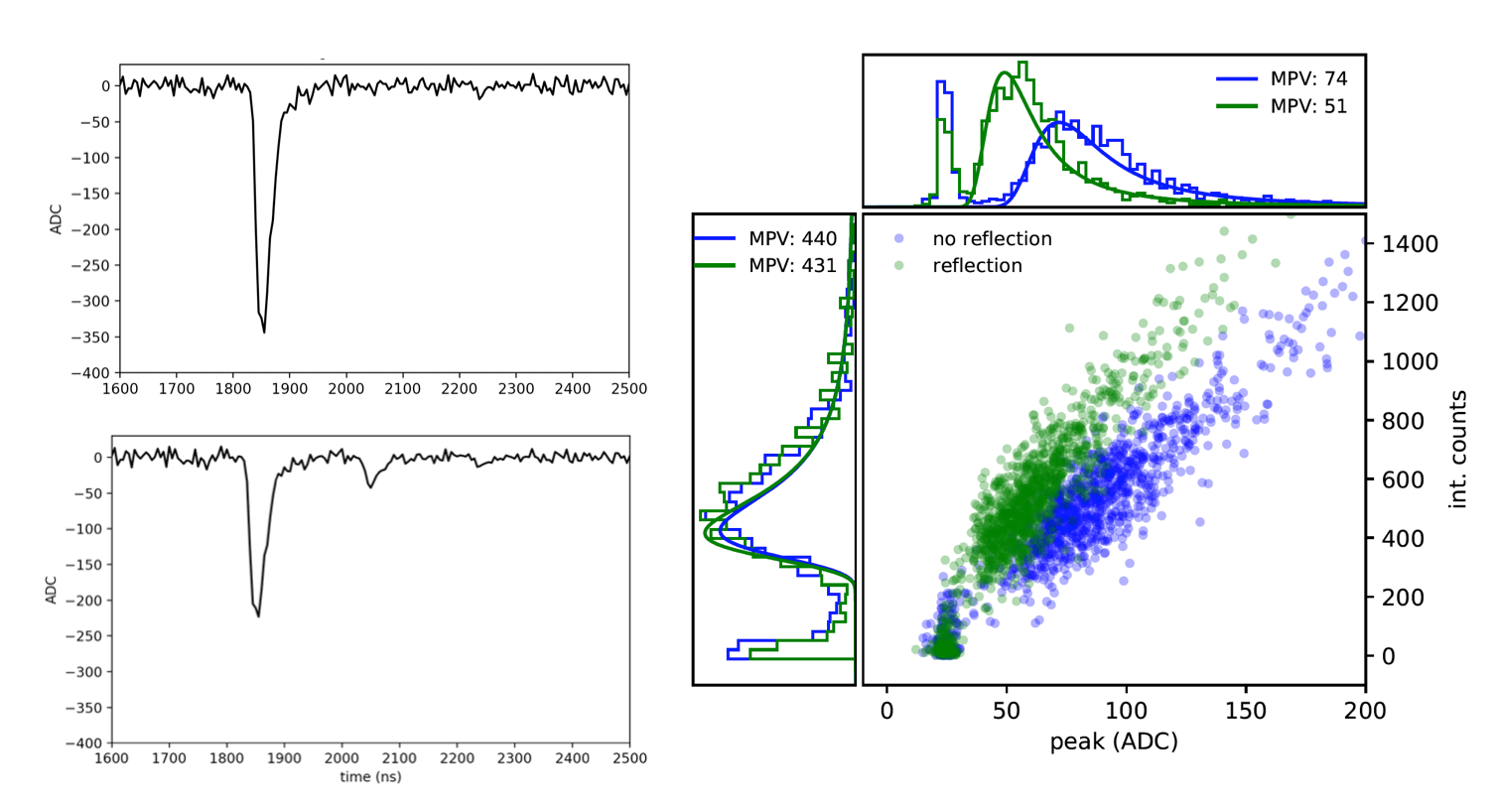}
\caption{\textbf{Left:} sample traces from one PMT.  Top: Single muon trace directly from one PMT.  Bottom: Trace from a single muon, with signals from two PMTs joined with a BNC splitter.  \textbf{Right:}  Scatter plot of integrated ADC counts (a proxy for deposited charge), against the peak ADC count in the trace, for data collected with reflections (in green) and without 
reflections (in blue).  Histograms of the distributions projected along each axis are shown on the side panels. }
\label{fig:reflections}
\end{figure}

The right panel shows a scatter plot of integrated ADC counts (a proxy for deposited charge), against the peak ADC count in the trace, for data collected with reflections (in green) and without 
reflections (in blue).  Histograms of the distributions projected along each axis are shown on the side panels.  Here, it can be seen that the integrated counts remain similar in both cases, which means the energy conversion from ADC counts to MeV is unaffected.  This is because the reflected pulse is also within the integration window.  We find that the integrated signal drops between 0 and 10\% across all scintillators when reflections are included.  This is added to the systematic uncertainties on the energy scale derived from particle data.  The peak ADC value drops by closer to 40\% however, which must be taken into consideration when handling trigger biases, as the trigger is based on signal over threshold.

\section{Conclusion}

LORA acts as a trigger for radio readout for cosmic-ray detection at the LOFAR telescope, and provides particle-based information about the primary cosmic ray used in event reconstruction.  An expansion is underway that doubles the number of scintillators and expands the effective area of the array.  This will increase the number of measured cosmic-ray events with a strong radio signal by 45\%, and also decrease the trigger bias against heavier elements prominent in low-energy events.
A study of the response of the scintillators has also been carried out.   Measurements at the KIT muon tower provided information about the spatial response of the detectors, which will be taken into consideration in event simulation and reconstruction.  System responses such as reflections in the cables can also be treated in data processing, to ensure that trigger biases are handled correctly.  The LORA expansion is well underway, with field deployment of the scintillators having taken place in the spring of 2019.  First data is expected in the coming months.

\begin{footnotesize}

\begin{tocless}
\bibliographylora{LOFARbib}
\bibliographystylelora{unsrt}
\end{tocless}

\end{footnotesize}

\newpage

\end{document}